\documentclass[apj]{emulateapj}

\usepackage{url,amsfonts,epsfig}
\usepackage{graphicx}
\usepackage{amsmath}
\usepackage{amssymb}
\usepackage[section]{placeins}
\usepackage{chngpage}
\usepackage{calc}
\usepackage{subfigure}
\usepackage{mathtools}
\usepackage{appendix}
\usepackage{natbib}
\usepackage{graphicx}
\makeatletter 
\renewcommand\@biblabel[1]{}

\shorttitle{Dynamical friction of massive black hole binaries}
\shortauthors{Dosopoulou and Antonini}
\begin{document}
\def\gap{\;\rlap{\lower 2.5pt
\hbox{$\sim$}}\raise 1.5pt\hbox{$>$}\;}
\def\lap{\;\rlap{\lower 2.5pt
 \hbox{$\sim$}}\raise 1.5pt\hbox{$<$}\;}

\newcommand\sbh{MBH}
\newcommand\NSC{NSC}
\title{Dynamical friction and the evolution of Supermassive Black hole Binaries: the final hundred-parsec problem }

\author{Fani Dosopoulou and Fabio Antonini}
\email{FaniDosopoulou2012@u.northwestern.edu,\\fabio.antonini@northwestern.edu}

\affil{Center for Interdisciplinary Exploration and Research in Astrophysics (CIERA) and Department of Physics and Astrophysics,
Northwestern University, Evanston, IL 60208}

\begin{abstract}
The  supermassive black holes originally in the nuclei of two merging galaxies will form a binary in the remnant core. The early evolution of the massive binary is driven by dynamical friction before the binary becomes 
``hard'' and eventually reaches coalescence through gravitational wave emission.
{ We consider the dynamical friction evolution of massive binaries consisting of a secondary hole orbiting inside a stellar cusp 
dominated by a more massive central black hole.}
In our treatment we include  the frictional force from stars moving faster than the inspiralling object which is neglected in the standard Chandrasekhar's treatment. We show that the binary eccentricity increases if the stellar cusp density profile rises less steeply than $\rho\propto r^{-2}$. In  cusps shallower than  $\rho\propto r^{-1}$ the frictional timescale can become very long due to the deficit 
of stars moving slower than the massive body. Although including the fast stars increases the decay rate, low mass-ratio binaries ($q\lesssim 10^{-3}$) in sufficiently massive galaxies have decay timescales longer than one Hubble time. During such minor mergers the secondary hole stalls on an eccentric orbit at a distance of order one tenth the influence radius of the primary hole (i.e., $\approx 10-100\rm pc$ for massive ellipticals). We calculate the expected number of stalled satellites as a function of the host galaxy mass, and show that the brightest cluster galaxies should have $\gtrsim 1$ of such satellites orbiting within their cores. Our results could provide an explanation to a number of observations, which include multiple nuclei in core ellipticals, off-center AGNs and eccentric nuclear disks.\\
 \end{abstract}


\section{Introduction}
 A massive object moving through a cluster of lighter stars suffers a net deceleration along the direction of 
its motion known as dynamical friction. 
Dynamical friction can be understood as the drag induced on a test particle by the overdensity (i.e., the gravitational wake) that is raised behind it by the deflection 
of stars \citep{DC:57,kln:72,MU:83,1986ApJ...300...93W}. 
Dynamical friction is one of the most fundamental 
processes in astrophysics and its understanding is arguably the most important contribution 
of Chandrasekhar to stellar dynamics \citep{1943ApJ....97..255C}.
Dynamical friction plays a key role in the evolution of 
supermassive black hole (SMBH) binaries \citep[e.g.,][]{2006ApJ...648..976M}, 
galaxies \citep[e.g.,][]{1999ApJ...515...50V}, star clusters \citep[e.g.,][]{2014ApJ...795..169A},
binary star cores in the common envelope phase of evolution \citep[e.g.,][]{1976IAUS...73...75P}, and 
 protoplanet migration \citep[e.g.,][]{1999ApJ...513..252O}. 

 Chandrasekhar describes dynamical friction
 as the systematic decelerating effect of the fluctuating field of force 
 acting on a star in motion.
By  assuming that the unperturbed motion of the test body
was linear and unaccelerated, and that the field-star distribution
was infinite and homogeneous spatially and isotropic in velocity
space, Chandrasekhar
  derived an explicit formula for the dynamical friction force \citep{1943ApJ....97..255C}
 \begin{equation}\label{ch43}
 \pmb{F}_{df}=-4\pi G^{2} \frac{\pmb{\upsilon}}{\upsilon^{3}}
(m+m_\star)  \ln{\Lambda} \int_{0}^{\upsilon}d\upsilon_{\star} 4\pi f(\upsilon_{\star})\upsilon_{\star}^{2},
  \end{equation}
  where $m_\star$ denotes the mass of the field stars,
  $m$  the mass of the test body, $v$ its velocity, $G$ the gravitational constant, $\ln \Lambda$ the Coulomb logarithm,
  and $f(v_\star) $  the stellar velocity distribution function.
 Clearly, Equation\ (\ref{ch43}) implies 
 that  only stars with velocity $v_\star <v$, i.e., that move \emph{ slower} than the test body, contribute to the decelerating force. 
 Although Equation\ (\ref{ch43})  has been shown to describe remarkably well a variety of systems 
 \citep[e.g.,][]{2003MNRAS.344...22S,2006MNRAS.372..174B}, 
 deviations from the standard theory are also known to exist \citep[e.g.,][]{TW1984,2006MNRAS.373.1451R,2008ApJ...678..780G,2012ApJ...745...83A,2014ApJ...785...51A,2016MNRAS.463..858P}.
 The motivation for the work described in this paper is the existence of physically interesting models of galactic  nuclei in which
 Equation\ (\ref{ch43}) 
leads to erroneous  results because the usual simplifying assumptions
that lead to neglect the contribution of  stars with large velocities ($v_\star >v$) appear to break down.

In this paper we present a comprehensive study of dynamical friction in the nuclei of galaxies 
containing a SMBH.
We derive the dynamical friction coefficients which describe the orbit-averaged time evolution 
of the  energy and angular momentum  of a test star near a SMBH.
We do this by using the proper field star
velocity distribution (as opposed to, say, a
Maxwellian), and including the contribution of the fast stars to the
frictional force that was ignored in previous treatments of this problem.
More specifically, analytical techniques and $N$-body simulations
 are used to test two  predictions that the standard Chandrasekhar's treatment
makes about the evolution 
of a massive body moving  near a SMBH.
Assuming that the density of field stars follows 
 $\rho (r)\sim r^{-\gamma}$, then Equation\  (\ref{ch43}) predicts that \citep[][]{2012ApJ...745...83A}:
 (i) the dynamical friction force goes to zero
 as $\gamma\rightarrow 1/2$, (ii) 
 the eccentricity is conserved for $\gamma=3/2$, while dynamical friction tends to circularize 
 orbits for 
$\gamma>3/2$ and make them more eccentric for $\gamma<3/2$. Compared to previous work \citep[e.g.][]{2012ApJ...745...83A}
we study the evolution of massive binaries in models with a wide
range of density profiles and binary mass ratios and consider for the first time the effect of the friction from fast stars
to the evolution of the binary eccentricity. Moreover, in our paper the binary evolution equations are first derived using a perturbation approach 
based on the varying conic method of Lagrange 
 \citep[e.g.,][]{2016ApJ...825...70D} and then orbit-averaged \citep[e.g.,][]{2016ApJ...825...71D}.

Predictions (i) and (ii) can be easily understood based on 
Equation\  (\ref{ch43}), and noting that 
the distribution function which corresponds to an isotropic and spherical cluster of stars near a SMBH is
 $f(E)\propto {|E|^{\gamma-3/2}/ \Gamma(\gamma-1/2)}$ [see also Equation (\ref{dis1}) below], with $E$ the orbital energy. The previous expression shows that as 
$\gamma \rightarrow 1/2$ all stars have zero energy with respect to the SMBH, i.e., 
the distribution function tends to a delta function which peaks at the local escape velocity.
Under these conditions, Equation\  (\ref{ch43})  implies that the 
dynamical friction force is zero, since there are no stars locally that move
slower than the test star. 
Point (ii) is also easily shown to be true.
For $\gamma= 3/2$, $f(E)=\rm const$. Equation\ (\ref{ch43}) in this case
shows that the dynamical friction force is independent of radius and reduces to a 
linear deceleration drag ${\bf F}_{df}=-\rm const \cdot {\bf v}$ along the orbit. This implies that the orbital eccentricity of a massive body
will remain unchanged during its motion. 
For steeper profiles, because the phase-space density is higher than average at periapsis
the additional drag there tends to circularize the orbit, while for shallower profiles the higher phase-space density at 
apoapsis tends to make the orbit more eccentric.

Our calculations   show that the evolution of the test mass can be significantly affected by 
 the frictional force produced by stars moving faster than its velocity.
 Adding this contribution leads to  a timescale for inspiral that, for $1/2 \lesssim \gamma\lesssim 1$, {can be  
 up to one order of magnitude shorter} than what predicted by 
Equation\  (\ref{ch43}).
The orbital eccentricity of the test mass is found to 
increase during the inspiral for all values of $\gamma$ less than $\approx 2$; for steeper
profiles the eccentricity decreases but only mildly before the secondary SMBH reaches
the center.

 Finally, we consider
the dynamical evolution  of SMBH binaries, the formation of which is believed
to be a generic product of galaxy mergers.
We explore the dependence of the lifetime of a SMBH binary on its total mass, mass ratio, and on the
density profile of the surrounding cusp. 
%We find that massive binaries with a low mass ratio orbiting within a  
%shallow density profile  cusp  ($\gamma <1$) 
%have dynamical friction
% timescales  that can largely exceed the Hubble time, leading 
% to a ``stalled'' satellite SMBH.
% Such nuclei appear to be common. For instance, 
%giant elliptical galaxies with flat inner cores have been observed for decades \citep[e.g.,][]{1966ApJ...143.1002K,1972IAUS...44...87K}. 
We calculate the expected  number of stalled satellites as a function of the host galaxy SMBH mass and find that the inner  cores  of massive galaxies like M87 
are likely hosts of stalled satellite SMBHs.  %Such satellites are predicted
%to orbit
%at a distance from the center of the main galaxy that is 
%of order the influence radius of the primary MBH.
The implications of our results are discussed in connection to a number of observations, which 
include off-center AGNs   \citep{2014ApJ...795..146L}, binary AGNs \citep{2006ApJ...646...49R},
double or multiple nuclei within core elliptical galaxies \citep{2016ApJ...829...81B,2016MNRAS.462.2847M} and eccentric nuclear disks \citep{1996ApJ...471L..79L}.

This paper is structured as follows. In Section 2 we introduce the general formalism we adopt
 to describe the orbital evolution of a massive binary moving inside a stellar cusp around a central SMBH, treating dynamical friction as a perturbation to the classic Kepler problem.  In Section 3 we compare the theoretical predictions for the binary evolution with the results of $N$-body simulations. In Section 4 we describe the different phases involved in the evolution of a SMBH binary and calculate the lifetime of a SMBH binary in early-type galaxies. In Section 5 we calculate the expected average number of stalled satellites in luminous galaxies as a function of the host galaxy 
SMBH mass, commenting on possible connections to observations. In Section 6 we present our conclusions.

\section{Analytical treatment}\label{analytical}

A binary system can be exposed to various  perturbations emerging from physical processes involved in the course of its evolution. Within the astrophysical context these processes are in principle dynamical processes in addition to the classic Newtonian gravity.

 In stellar binaries these processes include tidal forces and tidal friction, relativistic corrections, gravitational-wave emission, magnetic braking, mass-loss and mass-transfer interactions as well as many-body forces.  In massive binaries moving inside a stellar cusp, a fundamental perturbation is dynamical friction which is the deceleration drag experienced by the secondary massive body.

 Due to the various perturbations, each body in the binary is no longer moving in the actual Keplerian ellipse it would if no perturbations existed, but its physical orbit is slowly changing with time. The time evolution of the orbital elements can be described using the \emph{Varying-Conic method} advanced and completed by Lagrange. In this method the true physical orbit of the body is approximated by a family of evolving instantaneous ellipses that at each moment in time describe the ellipse the body would follow if the perturbation ceased instantaneously.

 In what follows we describe the general formalism of this method and we then apply this formalism to the orbital evolution of an inspiraling object inside a stellar cusp treating dynamical friction as a perturbation to the binary orbit.

\subsection{Varying-Conic method}

The general reduced two-body problem where Newtonian gravity is the only force acting on the two bodies in the system is described by the equation
\begin{align}\label{eq1}
\ddot{\bold{r}}=-\frac{GM}{r^{3}}\bold{r}
 \end{align}
where $G$ is the gravitational constant, $M$ is the total mass of the system and $r$ is the relative position between the two bodies.

Any dynamical interaction between the two bodies introduces an extra force to the binary which acts as a perturbation to the classic two-problem. Under the effect of a perturbing force $\bold{F}(\bold{r},\dot{\bold{r}})$ the equation of motion for the \emph{perturbed} two-body problem can be written as
\begin{align}\label{aa}
\ddot{\bold{r}}=-\frac{GM}{r^{3}}\bold{r}+\bold{F}(\bold{r},\dot{\bold{r}})
 \end{align}
where the perturbing force $\bold{F}(\bold{r},\dot{\bold{r}})$ depends in principle upon both the relative position $\bold{r}$ and velocity $\pmb{v}=\dot{\bold{r}}$.

 Equation $(\ref{aa})$ can be solved assuming that at each instant of time, the true orbit can be approximated by an instantaneous ellipse which is changing over time through its now \emph{time-dependent} orbital elements $C_{i}(t)$. Here $C_{i}=(a,e,i,\Omega,\omega,f)$ are namely the semi-major axis, eccentricity, inclination, longitude of the ascending node, argument of periapsis and true anomaly $f$. We also define $n=(GM/a^{3})^{1/2}$ as the mean motion.
 At each moment of time these orbital elements describe the orbit the body would follow if perturbations were to cease instantaneously. We refer to these elements as \emph{osculating} orbital elements.
 The time-evolution equations for the osculating orbital elements decomposed in the reference system $K_{R}({\bold{\hat{r}},\bold{\hat{\tau}},\bold{\hat{n}}})$, where the unit vector $\bold{\hat{r}}$ is along the relative position vector between the two bodies in the binary, become
\begin{align}
\frac{da}{dt}&=\frac{2}{n\sqrt{1-e^{2}}}\left[F_{r}e\sin f + F_{\tau} (1+e\cos f)\right]\label{ar}\\
\frac{de}{dt}&=\frac{\sqrt{1-e^{2}}}{na}\left[F_{r}\sin f + F_{\tau}\left(\cos f + \frac{e+\cos f}{1+ e\cos f}\right)\right]\label{er}\\
\frac{di}{dt}&=F_{n}\frac{\cos(f+\omega)\sqrt{1-e^{2}}}{na(1+e\cos f)}\label{ir}\\
\frac{d\Omega}{dt}&=\frac{1}{na^{2}\sqrt{1-e^{2}}\sin i}F_{n}\frac{a(1-e^{2})\sin (f+\omega)}{1+e\cos f}\label{Omegar}\\
\nonumber \frac{d\omega}{dt}&=\frac{\sqrt{1-e^{2}}}{nae}\left(-F_{r}\cos f + F_{\tau} \sin f \frac{2+e\cos f}{1+ e \cos f}\right) \\
&-\frac{\cos i}{na\sin i}F_{n}\frac{\sqrt{1-e^{2}}\sin (f+\omega)}{1+e\cos f}\label{omegar}\\
\frac{df}{dt}&=\frac{n (1+e\cos f)^{2}}{(1-e^{2})^{3/2}}-\frac{d\omega}{dt}-\cos{i}\frac{d\Omega}{dt}\label{sigmar} \ .
\end{align}

 Equations  $(\ref{ar})$ and $(\ref{er})$ indicate that for a perturbing force always vertical to the orbital plane, i.e., $F_{r}=F_{\tau}=0$ the semi-major axis and the eccentricity do not change while Equations  $(\ref{ir})$ and $(\ref{Omegar})$ show that the inclination $i$ and the longitude of the ascending node $\Omega$ evolve only for a non-zero vertical to the orbital plane component of the perturbing force, i.e., $F_{n} \neq 0$. On the contrary, from Equation  $(\ref{omegar})$ we see that the periapsis is precessing for any non-zero perturbing force, i.e., $\bold{F}\neq 0$.

  In the following section we apply the formalism described above to study the effect of dynamical friction on the orbital evolution of a test
   mass moving inside a cluster of stars around a central SMBH. In this case dynamical friction acts as a perturbing force on the evolution of the inspiraling object. We begin with a brief introduction to dynamical friction as described initially by \citet{1943ApJ....97..255C} and further studied by \citet{2012ApJ...745...83A} in the case of a test mass moving inside a cluster of stars around a 
more massive central SMBH.

\subsection{Dynamical friction}
\begin{figure}
  %\centering
    \includegraphics[width=0.5\textwidth]{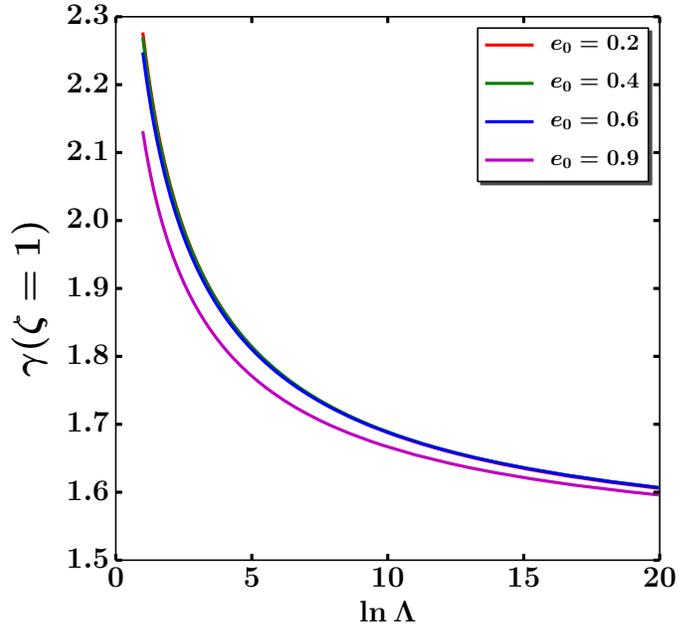}
      \caption{{Value of $\gamma$ for which $\zeta=1$ as a function of the Coulomb logarithm $\ln \Lambda$, and for different initial eccentricities $e_{0}$.
The eccentricity of the binary orbit increases with time for any $\gamma<\gamma(\zeta=1)$.
Note that if only the slow stars are included, $\gamma(\zeta=1)=1.5$ for any value of $\ln \Lambda$ and $e_{0}$.}}
      \label{ratioplot}
\end{figure}    

In what follows we consider 
the evolution of a  binary comprising a massive object moving near a SMBH of considerably larger mass and which sits at the center of a galaxy.
Gravitational interaction with stars results in loss of energy and angular momentum by the massive object.
We assume that the galaxy is spherically symmetric and isotropic and we describe it using a power-law stellar  density profile of the form $\rho (r)=\rho_{0}\left( r/r_{0}\right )^{-\gamma}$, where $r_{0}$ is a characteristic radius and $\gamma$ is the slope of the density profile. In what follows we set $r_{0}=r_{\rm infl}$, with $r_{\rm infl}$ 
 the radius containing a mass in stars twice  the mass of the central SMBH ($M_\bullet$), i.e.,  the  SMBH influence radius. 
  Unless otherwise specified, in what follows we use units such that  $M_\bullet=r_{\rm infl}=G=1$. %In these units, $\rho_0=(3-\gamma)/2\pi$.

 Assuming that the gravitational potential $\Phi$ is dominated by the central SMBH and neglecting the effect of the surrounding stars we can write $\Phi\approx-G M_{\bullet}/r$. Eddington's formula then uniquely leads to the following distribution function of the 
 field-star velocities \citep{2013degn.book.....M}
\begin{equation}
f(\upsilon_{\star})=\frac{\Gamma\left(\gamma+1\right)}{\Gamma\left(\gamma-\frac{1}{2}\right)}\:\frac{1}{2^{\gamma}\pi^{3/2}v_{c}^{2\gamma}}
\:\left(2v_{c}^{2}-\upsilon_{\star}^{2}\right)^{\gamma-3/2}\label{dis1}
\end{equation}
where $\upsilon_{\star}$ is the star velocity, $\upsilon_{c}=\sqrt{GM_{\bullet}/r}$ is the circular velocity and the normalization 
corresponds to  unit total number.

The general formula for the dynamical friction force including also the contribution from stars moving faster than the massive body is
\begin{align}
%\nonumber \pmb{F}_{df}\approx \: \:&\pmb{F}_{df}^{\left(\upsilon_{\star}<\upsilon\right)}+\pmb{F}_{df}^{\left(\upsilon_{\star}>\upsilon\right)}=\\
\begin{split}
 \pmb{F}_{df}\approx&-4\pi G^{2} m \rho(r) \frac{\pmb{\upsilon}}{\upsilon^{3}}
\times \left\lbrace \ln{\Lambda} \int_{0}^{\upsilon}d\upsilon_{\star} 4\pi f(\upsilon_{\star})\upsilon_{\star}^{2}\right.\\
 &\left. +\int_{\upsilon}^{\upsilon_{esc}}d\upsilon_{\star}4\pi
 f(\upsilon_{\star})\upsilon_{\star}^{2}\left[\ln\left({\frac{\upsilon_{\star}+\upsilon}{\upsilon_{\star}-\upsilon}}\right)-2\frac{\upsilon}{\upsilon_{\star}}\right] \right\rbrace\end{split}\label{df}
\end{align}
where $\pmb{\upsilon}$ is the velocity of the massive body and $m$ its mass. The quantity $\ln{\Lambda}$ is the Coulomb logarithm defined as
\begin{equation}\label{coulomb}
\ln{\Lambda}=\ln\left(\frac{b_{\rm max}}{b_{\rm min}}\right)\approx \ln\left(\frac{b_{\rm max}v_{c}^{2}}{Gm}\right)
\end{equation}
where $b_{\rm max}$ and $b_{\rm min}$ are the maximum and minimum impact parameters respectively.

 We can rewrite the dynamical friction force in a more compact form as 
\begin{align}
\bold{F}_{df}&=\kappa (r) \left[ \ln{\Lambda}\:\: \alpha(v) + \beta (v) + \delta (v) \right] \frac{\pmb{v}}{v^{3}}\label{dec}\\
\bold{F}_{df}&=\epsilon (r,v) \frac{\pmb{v}}{v^{3}}\label{df2}
\end{align}
where we defined 
\begin{align}
\kappa (r)&=-4\pi G^{2}\rho(r)m\label{int1}\\
\epsilon (r,v) &=\kappa (r) \left[ \ln{\Lambda}\:\: \alpha(v) + \beta (v) + \delta (v) \right]\label{int2}\\
\alpha (v)&=4\pi\int_{0}^{v} f(\upsilon_{\star}) \upsilon_{\star}^{2} d\upsilon_{\star}\label{int3}\\
\beta (v)&=4\pi\int_{v}^{v_{\rm esc}} f(\upsilon_{\star}) \upsilon_{\star}^{2} \left[\ln\left({\frac{\upsilon_{\star}+\upsilon}{\upsilon_{\star}-\upsilon}}\right)\right] d\upsilon_{\star}\label{int4}\\
\delta (v)&=4\pi v \int_{v}^{v_{\rm esc}} f(\upsilon_{\star}) (-2\upsilon_{\star})d\upsilon_{\star}.\label{int5}
\end{align}

 The distribution function $(\ref{dis1})$ can be rewritten as
\begin{equation}
f(\upsilon_{\star})=\frac{\Gamma\left(\gamma+1\right)}{\Gamma\left(\gamma-\frac{1}{2}\right)}\:\frac{1}{2^{\gamma}\pi^{3/2}v_{c}^{3}}\left[2-x^{2}\right]^{b}\label{dis2}
\end{equation}
where we defined $x=\upsilon_{\star}/v_{c}$ and $b=\gamma - 3/2$. Integrals $(\ref{int3})$ and $(\ref{int5})$ have an analytic form while integral $(\ref{int4})$ demands numerical manipulation. Using Equation $(\ref{dis2})$ we can rewrite the above integrals in a dimensionless form as
\begin{align}
\nonumber \alpha (\xi)&=\frac{\Gamma\left(\gamma+1\right)}{\Gamma\left(\gamma-\frac{1}{2}\right)}\:\frac{4}{3}\pi^{-1/2}2^{b-\gamma} \xi^{3} \\ &\:\:\:\:\:\: \times _{2}F_{1} \left[ 3/2,-b,5/2,\xi^{2}/2\right]\label{int3-2}\\
\beta (\xi)&=\frac{\Gamma\left(\gamma+1\right)}{\Gamma\left(\gamma-\frac{1}{2}\right)}4\pi^{-1/2}2^{-\gamma}\\
& \:\:\:\:\: \times \int_{\xi}^{1.4}  x^{2}(2-x^{2})^{b} \ln\left({\frac{x+\xi}{x-\xi}}\right) dx\label{int4-2}\\
\delta (\xi)&=\frac{\Gamma\left(\gamma+1\right)}{\Gamma\left(\gamma-\frac{1}{2}\right)} 8\pi^{-1/2}\frac{2^{-\gamma -1}}{b+1}\xi\\
& \:\:\:\:\: \times \left[ 0.04^{b+1}- (2-\xi^{2})^{b+1}\right] \label{int5-2}
\end{align}
where $\xi=v/v_{c}$  and we made use of that fact that $v_{\rm esc}/v_{c}=\sqrt{2}$.

The orbital velocity and dynamical friction force $(\ref{df2})$ decomposed in the reference system $K_{R}({\bold{\hat{r}},\bold{\hat{\tau}},\bold{\hat{n}}})$ mentioned above can be written as
\begin{align}
\pmb{\upsilon}&=v_{r}\bold{\hat{r}}+v_{\tau}\bold{\hat{\tau}}+v_{n}\bold{\hat{z}}\label{dec1}\\
\bold{F}_{df}&=F_{df,r}\bold{\hat{r}}+F_{df,\tau}\bold{\hat{\tau}}+F_{df,n}\bold{\hat{z}}\label{dec2}\ .
\end{align}

Substituting Equation (\ref{dec1}) into Equation (\ref{df2}) we have  $F_{df,r}=\epsilon (r,v) \frac{v_{r}}{v^{3}}$,  $F_{df,\tau}=\epsilon (r,v) \frac{v_{\tau}}{v^{3}}$ and $F_{df,n}=v_{n}=0$.

Using the dynamical friction components derived above and Equations $(\ref{ar})-(\ref{sigmar})$ the \emph{osculating orbital element time-evolution equations of an inspiraling massive body due to dynamical friction} are
\begin{align} 
\frac{da}{dt}&=\frac{2\epsilon (r,v)}{n^{3}a^{2}}\frac{(1-e^{2})^{1/2}}{(1+e^{2}+2e\cos{f})^{1/2}}\label{da}\\
\frac{de}{dt}&=\frac{2\epsilon (r,v)}{n^{3}a^{3}}(1-e^{2})^{3/2}\frac{e+\cos{f}}{(1+e^{2}+2e\cos{f})^{3/2}}\label{de}\\
\frac{di}{dt}&=F_{df,n}\frac{\cos(f+\omega)\sqrt{1-e^{2}}}{na(1+e\cos f)}=0\label{di}\\
\frac{d\Omega}{dt}&=\frac{1}{na^{2}\sqrt{1-e^{2}}\sin i}F_{df,n}\frac{\cos(f+\omega)\sqrt{1-e^{2}}}{na(1+e\cos f)}=0\label{dOmega}\\
\frac{d\omega}{dt}&=\frac{2\epsilon (r,v)}{n^{3}a^{3}}\frac{(1-e^{2})^{3/2}}{(1+e^{2}+2e\cos{f})^{3/2}}\frac{\sin f}{e}\label{domega}\\
\frac{df}{dt}&=\frac{n (1+e\cos f)^{2}}{(1-e^{2})^{3/2}}-\frac{d\omega}{dt}\label{dsigma}.
\end{align}
Equations $(\ref{da})-(\ref{dsigma})$ verify the aforementioned comment that in the absence of a vertical to the orbital plane component of the perturbing force, which is true in the case of dynamical friction ($F_{df,n}=0$), the inclination $i$ and the longitude of the ascending node $\Omega$ remain constant in the absence of other perturbing forces. Although according to Equation $(\ref{domega})$ dynamical friction can in principle induce a precession to the orbit, this precession has a negligible effect on the evolution of $e$ and $a$.
% Dynamical friction though affects strongly the semi-major axis and the eccentricity of the inspiraling body.

%\par Equations $(\ref{da})-(\ref{dsigma})$ are in principle phase-dependent. The long dynamical friction timescale indicates that the orbit-average of the above equations is adequate to describe the time evolution of the SMBH binary. In Section 5 using the orbit-averaging technique we derive the secular semi-major axis and eccentricity evolution of the massive binary.

\subsection{Eccentricity evolution}
We focus  here on how dynamical friction affects the binary eccentricity. We begin by investigating qualitatively the expected eccentricity evolution using a simplified physical picture of the problem. This picture focuses on the eccentricity changes near periapsis and apoapsis. This analysis is
 useful to understand the link between the expected eccentricity evolution of the system and the physical origin of the dynamical friction force.
The time-evolution of the eccentricity vector $\bold{e}$ induced by a perturbing force $\bold{F}$ (in our case $\bold{F}=\bold{F}_{df}$) is given by
\begin{equation}\label{eccchange}
\bold{\dot{e}}=\frac{1}{GM}\left(\bold{F}_{df}\times \bold{h} + \pmb{\upsilon}\times \bold{\dot{h}} \right)
\end{equation}
where $\bold{h}=\bold{r}\times \pmb{\upsilon}$ is the angular momentum per unit reduced mass $\mu=M_\bullet m/(M_\bullet+m)$ and the dot indicates a derivative with respect to time. In the absence of other perturbing forces in the binary, the angular momentum changes are only due to dynamical friction. 

 The dynamical friction force given by Equation $(\ref{df})$ is a decelerating drag-force (i.e., $\bold{F}_{df}=-g(\upsilon,r)\pmb{\upsilon}$ where $g(\upsilon,r)$ is a function of the massive body position and velocity) always acting in the opposite to the motion of the body direction. In addition, the time-evolution of the angular momentum vector is given by $\bold{\dot{h}}=\bold{r}\times  \bold{F}_{df} \Rightarrow \bold{\dot{h}}= -g(\upsilon,r)\bold{r}\times \pmb{\upsilon}$. This leads to $\bold{\dot{e}}=-2 g(\upsilon,r) \left[\upsilon^{2}\bold{r}- (\pmb{\upsilon}\cdot \bold{r}) \:\: \pmb{\upsilon}\right]$. The eccentricity vector $\bold{e}$ is defined as always pointing in the direction of periapsis. This indicates that the eccentricity rate induced at periapsis is $\dot{e}_{p}=\bold{\dot{e}}\cdot \hat{\bold{r}} <0$ tending to decrease the eccentricity and circularize the orbit, while at apoapsis $\dot{e}_{a}=\bold{\dot{e}}\cdot \hat{\bold{r}} >0$ tending to increase the eccentricity (note that $\pmb{\upsilon}\cdot \bold{r}=0$ at both periapsis and apoapsis). In addition, Equation (\ref{eccchange}) shows that for a massive body in an elliptical orbit the eccentricity decrease rate due to dynamical friction is maximized at periapsis ($f=0$) while the eccentricity increase rate is maximized at apoapsis ($f=\pi$). This implies that the eccentricity evolution depends on the relative time the massive body spends near periapsis and apoapsis along the orbit. Using this simplified picture we calculate below the expected eccentricity evolution of the binary orbit.

\begin{figure}
  \centering
    \includegraphics[width=0.5\textwidth]{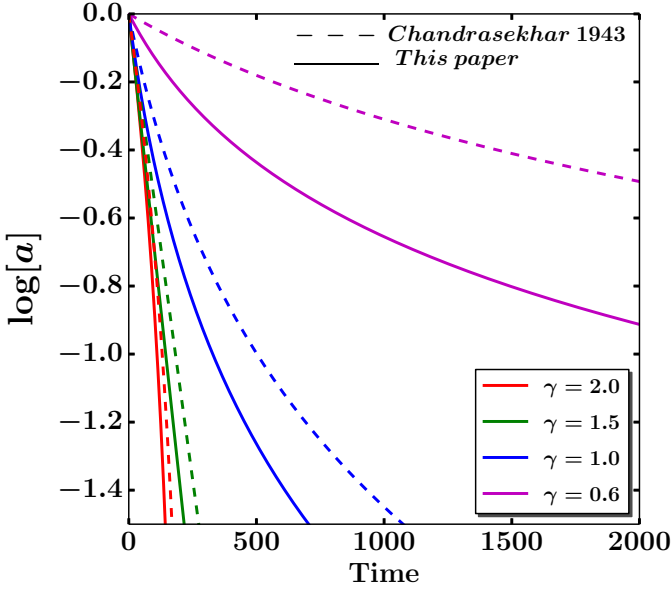}
      \caption{Secular evolution of the semi-major axis for a massive binary with $q=10^{-3}$
      and for different slopes $\gamma$ of the density profile. Dashed lines include only the slow stars contribution to dynamical friction \citep{1943ApJ....97..255C} while solid lines include also the contribution from fast stars. In shallow stellar cusps with $\gamma <1$ the dynamical friction timescale becomes long and the orbital decay is slow. Adding the contribution from the fast stars increases the orbital decay rate. In this calculation we set $\ln\Lambda = 6.0$ and $e_{0}=0.1$.}
      \label{aevolution}
\end{figure}

%We can  predict the expected eccentricity evolution, investigating the eccentricity changes induced to the binary orbit 
%near periapsis and apoapsis. 
Equation $(\ref{de})$ gives the instantaneous change of the eccentricity 
\begin{align}
\left( \frac{de}{dt} \right) _{p}&=\frac{2(1-e^{2})^{3/2}}{n^{3}a^{3}}\frac{\epsilon_{p}}{(1+e)^{2}}\label{dep}\\
\left( \frac{de}{dt} \right) _{a}&=-\frac{2(1-e^{2})^{3/2}}{n^{3}a^{3}}\frac{\epsilon_{a}}{(1-e)^{2}},\label{dea}
\end{align}
with $\epsilon_{p}$ and $\epsilon_{a}$ the value of $\epsilon(r,v)$ calculated at periapsis and apoapsis respectively.
 The time spent by the massive body near periapsis $(\Delta \tau_{p})$ or apoapsis $(\Delta \tau_{a})$ 
are proportional to
%is ill-defined due to the vague choice of a proportionality constant. These quantities are proportional to
\begin{align}
\Delta \tau_{p}\propto \frac{1}{\upsilon_{p}} \sim \sqrt{\frac{1-e}{1+e}}\label{dtp}\\
\Delta \tau_{a}\propto \frac{1}{\upsilon_{a}} \sim \sqrt{\frac{1+e}{1-e}}\label{dta}.
\end{align} 
The expected increase or decrease of the eccentricity can then be determined by the ratio
\begin{equation} 
\zeta\equiv { \Delta e_p\over \Delta e_a}=\frac{(de/dt)_{p}\: \Delta \tau_{p}}{(de/dt)_{a} \: \Delta \tau_{a}}\label{zeta}\ ,
\end{equation}
where $\Delta e_p$ and $\Delta e_a$ are the induced eccentricity changes
near periapsis and apoapsis respectively.
If $|\zeta |=1$ the eccentricity remains constant, if $|\zeta | > 1$ the contribution near periapsis dominates and the eccentricity decreases while if  $|\zeta | < 1$ the apoapsis contribution dominates and the eccentricity increases. 

%For the purpose of a qualitative study of the eccentricity evolution we can proceed assuming that throughout the whole path near periapsis or apoapsis the time derivative of eccentricity is constant and equal to the maximum value at periapsis and apoapsis respectively. 

We note that Equations (\ref{dep}) and (\ref{dea}) reduce to analytic expressions if we take into account only stars moving slower than the massive body. Under this consideration we have $\beta(\xi)=\gamma(\xi)=0$ and Equations (\ref{int2}) and $(\ref{int3})$ lead to
\begin{align}
\epsilon_{p}&\propto (1-e)^{-\gamma}{(1+e)^{3/2}} \:\:\:{}_{2}F_{1}^{p}\label{ep}\\
\epsilon_{a}&\propto (1+e)^{-\gamma}{(1-e)^{3/2}} \:\:\:{}_{2}F_{1}^{a}.\label{ea}
\end{align}
 Substituting Equations $(\ref{ep})$ and $(\ref{ea})$ into Equations (\ref{dep}) and (\ref{dea}) we have
\begin{align}
\left( \frac{de}{dt} \right) _{p}&\propto -\frac{(1-e)^{-\gamma}}{(1+e)^{1/2}} \:\:\:{}_{2}F_{1}^{p}\label{depp}\\
\left( \frac{de}{dt} \right) _{a}&\propto \frac{(1+e)^{-\gamma}}{(1-e)^{1/2}} \:\:\:{}_{2}F_{1}^{a}\ ,\label{deaa}
\end{align}
where the hyper-geometric function $_{2}F_{1}$ is always positive for all $0<e<1$. The different sign in Equations $(\ref{depp})$ and $(\ref{deaa})$ confirms the fact that  dynamical friction tends to circularize the orbit at periapsis ($\dot{e}_{p} <0$) while at apoapsis tends to increase it ($\dot{e}_{a} >0$).
Combining Equations $(\ref{dtp})$, $(\ref{dta})$, $(\ref{depp})$ and $(\ref{deaa})$ the ratio (\ref{zeta}) is simplified to
\begin{equation}
|\zeta |=\left(\frac{1-e}{1+e}\right)^{3/2-\gamma} \frac{_{2}F_{1}^{p}}{_{2}F_{1}^{a}}.\label{absz}
\end{equation}
As expected, for $\gamma=3/2$ we find $|\zeta |=1$ and the eccentricity remains constant;
for  $\gamma > 3/2 $ {the first term on the right-hand side of Equation $(\ref{absz})$ is $>1$ and 
$_{2}F_{1}^{p} > _{2}F_{1}^{a} $. In this case  Equation $(\ref{absz})$ leads to $|\zeta | > 1$. On the other hand, for $\gamma < 3/2 $ we have $_{2}F_{1}^{p} < _{2}F_{1}^{a} $ and $|\zeta | < 1$.}

Figure \ref{ratioplot} demonstrates the expected eccentricity evolution. 
 When the contribution of fast stars is included, the critical value of $\gamma$  below which 
 the eccentricity increases  is no longer $3/2$ but  $\approx2$. This critical value of $\gamma$  is found to depend slightly on the initial eccentricity 
 {while a smaller $\ln \Lambda$ increases the parameter space for which the binary eccentricity increases.}
 The fact that the critical value of $\gamma$ in this case is greater than $3/2$ reflects the fact that the relative contribution of the 
 fast stars is larger near apoapsis where the massive body is moving slower than the local circular velocity. This results in an enhanced drag force at apoapsis and higher eccentricties.

\subsection{Secular evolution}

\begin{figure}
  \centering
   \includegraphics[angle=0,width=3.2in]{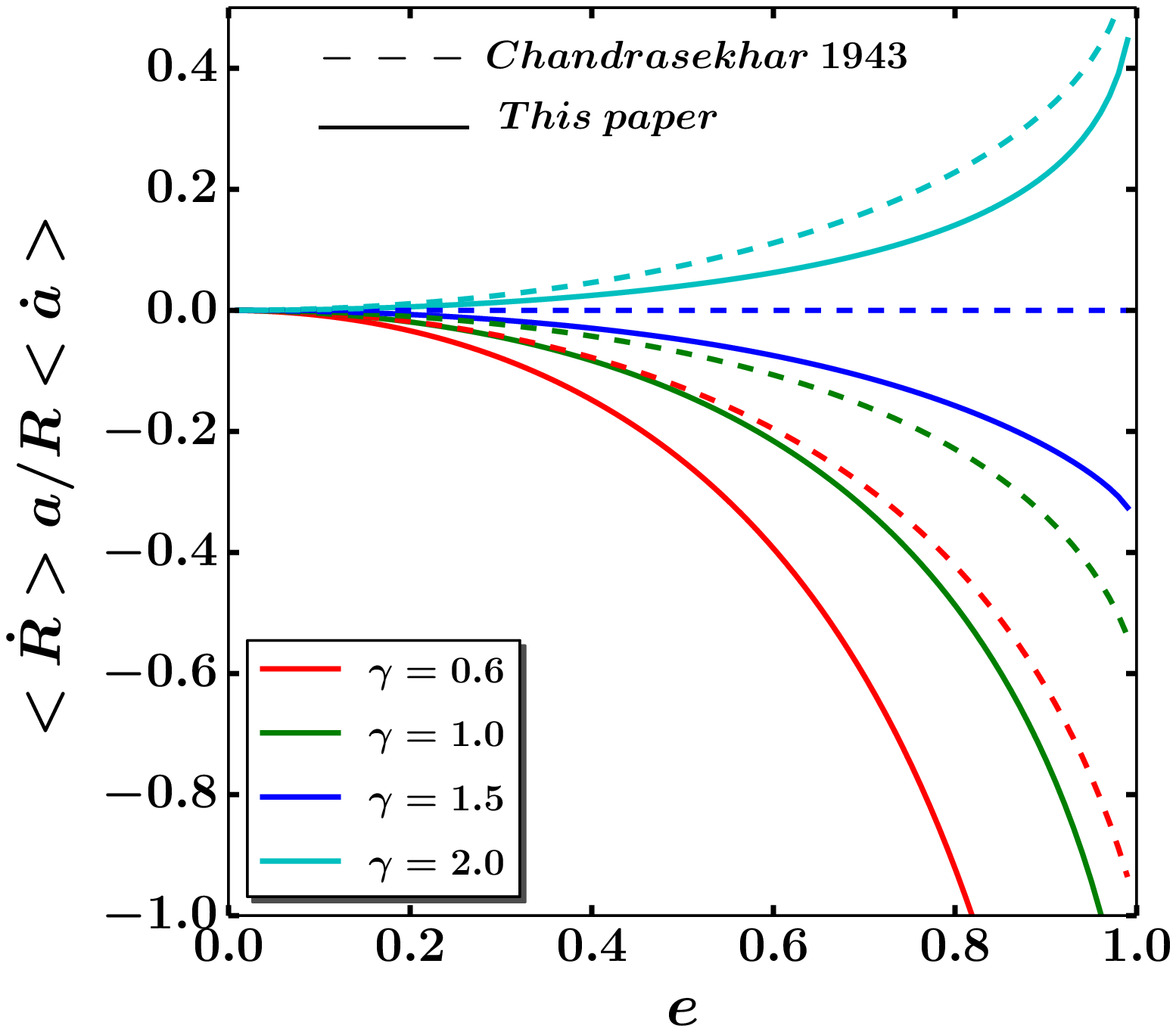} 
   \includegraphics[angle=0,width=3.2in]{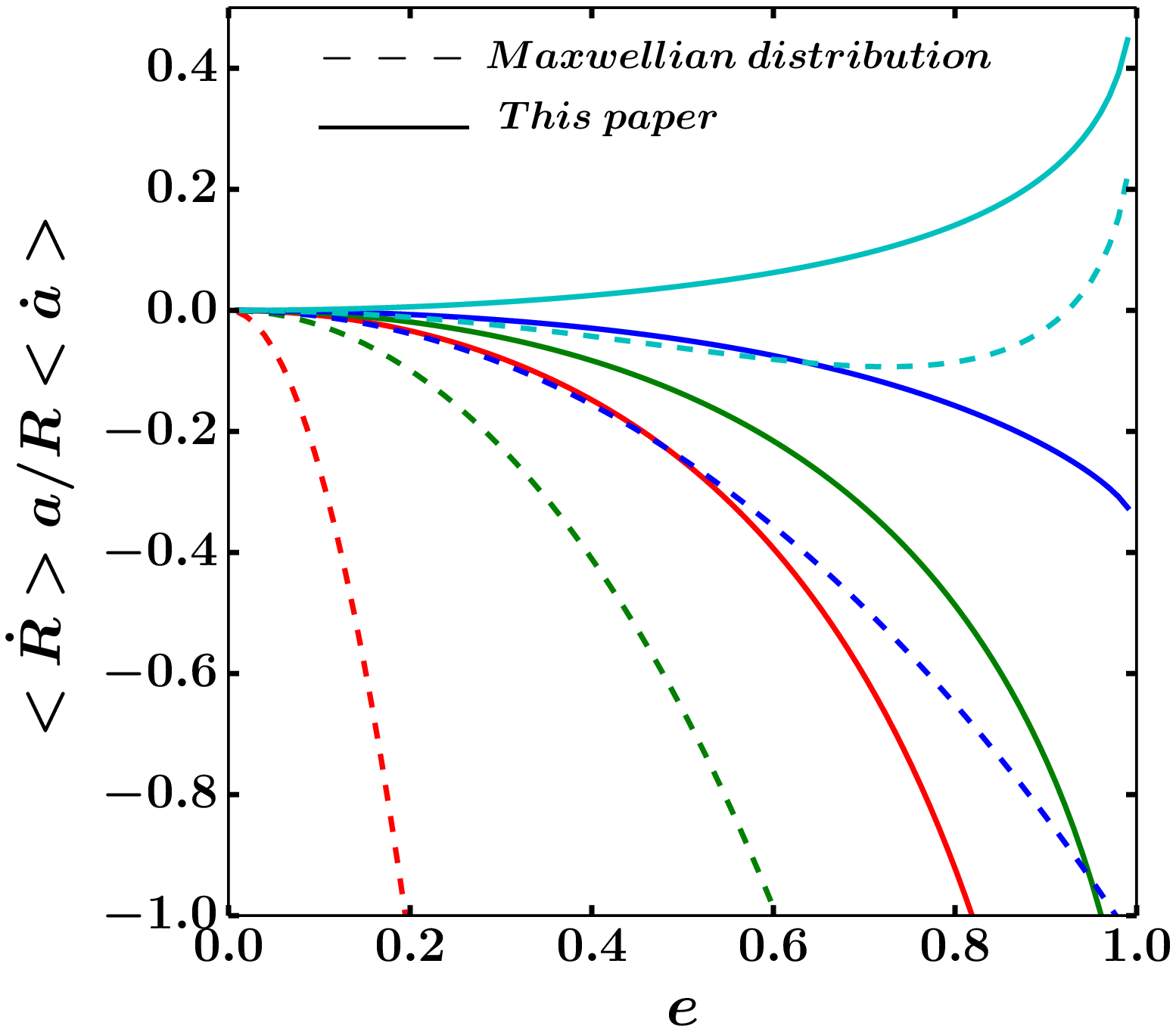}
      \caption{The average fractional change in the quantity $R=1-e^{2}$ by the time the massive body reaches the center as a function of the initial eccentricity. Solid lines include the contribution from the fast stars. The  dashed lines in the upper panel include only the contribution from stars moving slower than the test mass (e.g., Figure 8.18 in \citet{2013degn.book.....M}), while in the lower panel they are obtained by using a Maxwellian stellar velocity distribution. The eccentricity always increases unless $\gamma \sim 2$, and it is always higher when including the fast stars. Assuming a  Maxwellian velocity distribution appears to
     be a poor approximation for any value of $\gamma$.}
     \label{eevolution}
\end{figure}

The orbital element time-evolution Equations $(\ref{da})-(\ref{dsigma})$ of a perturbed binary are in principle phase-dependent and undergo physical oscillations with the orbital period. For a perturbation with a long timescale compared to the orbital period these oscillations can be smoothed out adopting orbit-averaging techniques. The orbit-averaged orbital element time-evolution equations then describe the secular evolution of the system.
\begin{figure*}
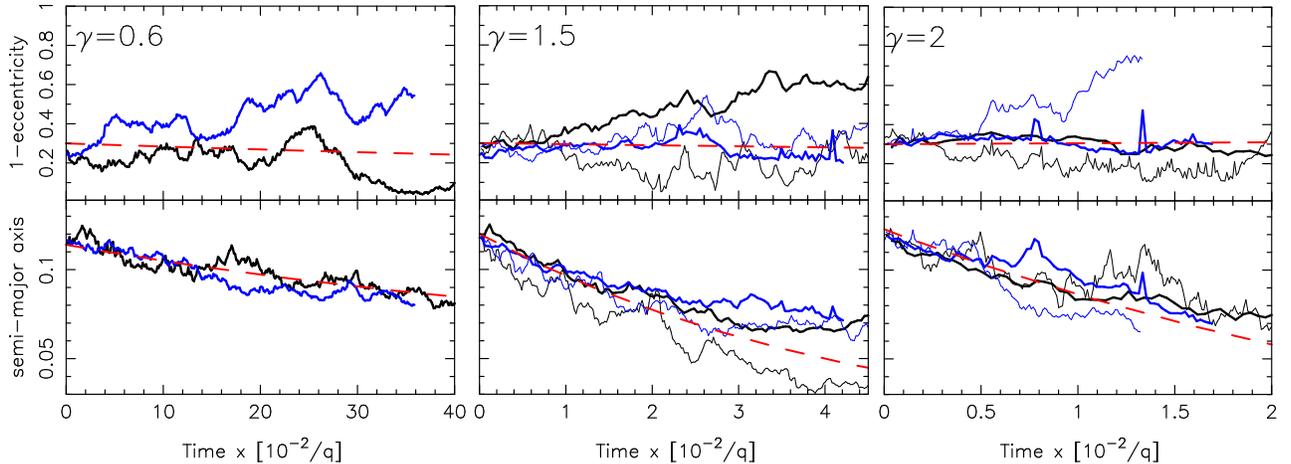

\begin{center}
 \includegraphics[angle=0,width=2.4in]{plot05f.eps} 
   \includegraphics[angle=0,width=2.08in]{plot15f.eps}
      \includegraphics[angle=0,width=2.09in]{plot2f.eps}
  \caption{Evolution of semi-major axis and eccentricity of the massive binary in the $N$-body simulations.  Results for  $m_\star=(5\times10^{-6}$ and $ 1\times10^{-5})$  are given by the black and blue line respectively. Line thickness increases with the mass ratio of the binary
for which we considered the two values: $q=(2\times10^{-4};\ 5\times10^{-5})$.
Simulations shown good agreement with the theoretical prediction (dashed red lines), which was obtained from
 Equation $(\ref{avinte})$  with  $\ln \Lambda\approx \ln \left( 1/10^{-4}\right)$.}\label{N-body0}
\end{center}
\end{figure*}
In order to orbit-average a quantity along the orbit we need to know how the true anomaly is changing over time. This is described by Equation $(\ref{sigmar})$ which shows that apart from the unperturbed Keplerian evolution described by the first term on the right-hand side of this equation the true anomaly can also evolve due to possible precessions and specifically the periapsis precession $\dot{\omega}$ and the longitude of the ascending node precession $\dot{\Omega}$. 
Given that  due to dynamical friction we have $\dot{\omega} << 1$ and that  $\dot{\Omega}=0$, we can compute the secular evolution of the orbital elements neglecting the second term in Equation $(\ref{sigmar})$ and use
\begin{equation}
 df=n\frac{(1+e\cos f)^{2}}{(1-e^{2})^{3/2}}dt\label{trued}
\end{equation} 
when integrating Equations (\ref{da}) and (\ref{de}) over the orbit. In order to incorporate higher-order effects in the time-evolution equations we have to proceed one step further and include all the terms in Equation $(\ref{sigmar})$ when applying the orbit-averaging process. In this paper, we derive first-order secular time-evolution equations where the general orbit-averaging rule for the phase-dependent quantity $(...)$ is defined by the  integral
\begin{equation}
\left\langle\: (...) \:\right\rangle=\frac{(1-e^{2})^{3/2}}{2\pi}
\int_{0}^{2\pi}\: (...) \: \frac{df}{(1+e\:cos f)^{2}}\label{avint} \ .
\end{equation}
%where we used equation (\ref{trued}) and the generally phase-dependent quantity to be orbit-averaged is depicted with the three dots. %In most cases, this makes the above integral tedious and not easy to resolve into a simple analytical form.
 Under these considerations, the secular time-evolution equations for the semi-major axis and the eccentricity of the massive body can be written as
\begin{align}
\left\langle\: \frac{da}{dt} \:\right\rangle=&\frac{2(1-e^{2})^{2}}{\pi n^{3}a^{2}}
\int_{0}^{2\pi}\:\frac{(1+e\cos f)^{-2}\epsilon (r,v)}{(1+e^{2}+2e\cos{f})^{1/2}}df\label{avinta}\\
\nonumber \left\langle\: \frac{de}{dt} \:\right\rangle=&\frac{2(1-e^{2})^{3}}{\pi n^{3}a^{3}}\\
&\times \int_{0}^{2\pi}\:\frac{(e+\cos f)\epsilon (r,v)}{(1+e^{2}+2e\cos{f})^{3/2}(1+e\cos f)^{2}}df.\label{avinte}
\end{align}
Making use of Equation $(\ref{avinta})$ we plot in Figure \ref{aevolution} the secular semi-major axis evolution of an inspiraling body with mass ratio $q=m/M_{\bullet}=10^{-3}$. 
We see that for $\gamma <1$ the dynamical friction timescale becomes much  longer, although 
the contribution to dynamical fiction from the fast stars always increases the orbital decay rate. We discuss the implications of the long dynamical friction timescale for SMBH binaries in early-type galaxies in Section \ref{formation} and \ref{stalledsection} below.

The secular evolution of the binary eccentricity is described by Equation $(\ref{avinte})$. Using this equation we plot in Figure \ref{eevolution} the average fractional change in the quantity $R=1-e^{2}$ in one orbital decay time, i.e., $|a/\dot{a}|$. We plot this change as a function of the initial  eccentricity. %For high eccentricities $e\sim 1$ we have $\Delta R/R \approx \Delta (1-e) /(1-e)$ and we can interpret this change as the average change in SMBH binary eccentricity.

In the upper panel of Figure \ref{eevolution} we see that adding the fast stars contribution  makes the eccentricity higher compared to the case where only the slow stars were taken into account. The results shown in Figure 3 clearly imply that the change in eccentricity can be of order unity if $\gamma <1$.

In the lower panel of Figure \ref{eevolution} we compare the results of Equation $(\ref{avinte})$ with the
eccentricity change predicted assuming that the stellar velocities follow a Maxwellian distribution. As expected, the use of a Maxwellian velocity distribution is inadequate to describe the evolution of the binary. More specifically,  adopting a  Maxwellian distribution always leads to a shorter timescale for the eccentricity evolution compared to what we obtain by  using the distribution function $(\ref{dis1})$.
%\par In the case of SMBH binaries the study of the eccentricity evolution is important since mergers of eccentric massive black hole binaries produce strong gravitational waves (GW) and high eccentricity gravitational wave inspirals are characterized by unique electromagnetic (EM) and GW observables that can be detected by a space-borne instrument like LISA. Also, eccentricity affects strongly the hardening and gravitational wave emission timescales which are shorter for higher eccentricities. This means an increase in the eccentricity makes massive binaries coalesce faster.

\section{$N$-body treatment}\label{N-body}

Chandrasekhar formulated his theory assuming 
an infinite and homogeneous field of stars and
that  the unperturbed field star trajectories are straight lines.
Any of these assumptions represents a simplification of
the real physical system; 
both the test and field particles for example move on ellipses, not straight lines, around the central SMBH.
Thus, it is not 
clear whether Equations $(\ref{avinta})$ and $(\ref{avinte})$
can accurately describe the dynamical evolution of the massive binary due to dynamical friction.
\citet{2012ApJ...745...83A} showed that Chandrasekhar's theory 
reproduces remarkably well the real decay rate of a massive object into 
a shallow density profile model, however only when
the contribution of the fast stars is included in evaluating the frictional drag. 
Whether  the theory also reproduces the evolution
of the binary eccentricity as well as
the orbital decay for a range of density profile slopes remains to be shown.
Here we  use $N$-body simulations
to  investigate the evolution of a massive binary 
starting with various initial eccentricities  and  density profile models.

In order 
to validate Chandrasekhar's treatment in the systems we considered,
the  results of the simulations are compared to theoretical predictions
 based on Equations $(\ref{avinta})$ and $(\ref{avinte})$. 
In order to make such comparison, in 
this section we adopt $\ln \Lambda\approx \ln (M_\bullet/m)$ which gives values consistent with those
 found in former studies \citep{2012ApJ...745...83A,2003MNRAS.344...22S} and can be derived 
 analytically from Equation (\ref{coulomb})
if one identifies $b_{\rm max}$
with the local scale length determined by the density gradient $\rho/|\nabla \rho|$ \citep{2011MNRAS.411..653J}.

Before we discuss the results of our $N$-body integrations
 we  introduce here two critical values of the binary separation which 
will turn out to be crucial for the correct interpretation of our models.
 Dynamical friction is expected to only affect the evolution of the massive binary 
until its separation reaches
the semi-major axis of a ``hard'' binary \footnote{Defined as a binary that ejects passing stars at typical velocities greater than the escape velocity from the nucleus.} , which is often expressed as
 \citep{2013degn.book.....M}
\begin{equation}\label{ah}
a_{\rm h} \approx 36 {q\over (1+q)^2} 
{M_\bullet+m\over 3\times 10^9M_\odot} \left( \sigma\over 300\ \rm km\ s^{-1}\right)^{-2} \rm pc\ .
% \frac{q}{4(1+q)^2}r_{\rm infl}\ ,
\end{equation} 
with \citep{2005PhR...419...65A}
\begin{equation}
\sigma^2 = {1\over 1+\gamma }{G M_\bullet\over r}
\end{equation} 
the stellar velocity dispersion of the primary galaxy.
At  $a_{\rm h}$ the evolution of the massive binary ceases to be driven by dynamical friction, and 
its  semimajor axis shrinks as the two massive objects interact with stars 
and eject them from the nucleus via gravitational slingshots.
Even before the binary reaches $a_{\rm h}$, our analytical treatment is expected to breakdown as
 the binary separation becomes smaller than the radius 
 containing a mass 
equal to the mass of the inspiraling body
\begin{equation}\label{acrit}
a_{\rm crit}\equiv r(M_m=m)=r_{\rm infl}\left( m\over 2M_\bullet \right)^{1\over 3-\gamma} \ ,
\end{equation}
with $M_m(r)$ the mass in stars within a sphere of radius $r$ from the primary SMBH.
At $a\lesssim a_{\rm crit}$ the 
analytical treatment breaks down as
 the star distribution in the cluster starts to be significantly affected by the 
motion of the massive intruder  \citep[e.g.,][]{2006MNRAS.372..174B,2007ApJ...656..879M}.
%The values of $a_{\rm h}$ and $a_{\rm crit}$ for each of our $N$-body models 
%are given in Figure\ \ref{N-body2} and\ \ref{N-body1}.
\begin{figure*}
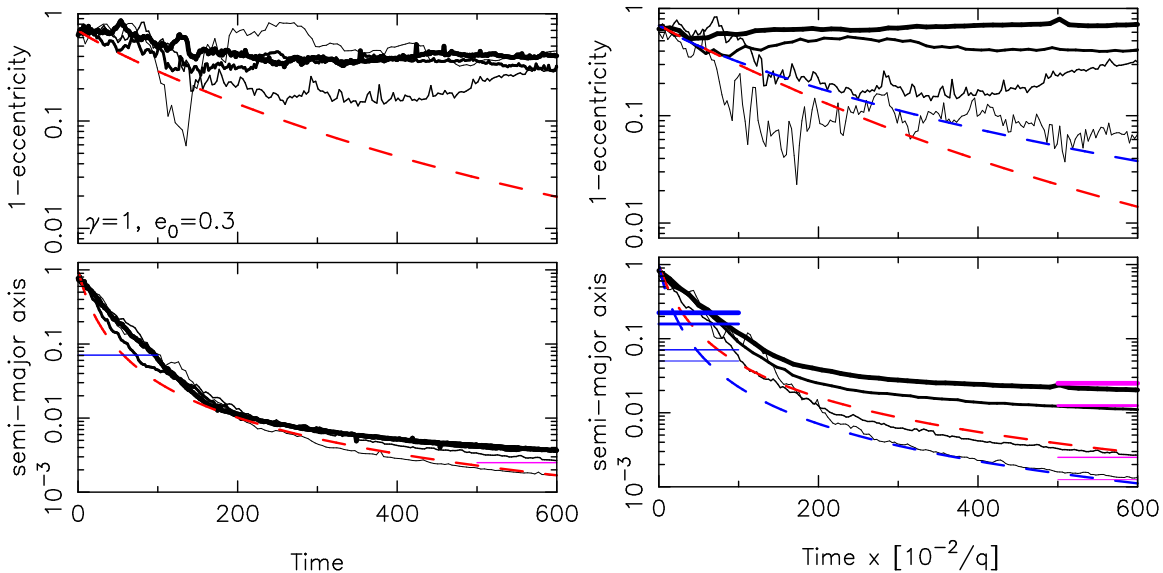

\begin{center}
 \includegraphics[angle=0,width=3in]{plotN.eps} 
 \includegraphics[angle=0,width=3in]{plotM.eps} 
  \caption{\textbf{Left panel:} Evolution of semi-major axis and eccentricity for  binaries with $q=10^{-2}$ and for  $m_{\star}=(4\times 10^{-4},2\times 10^{-4},10^{-4},5\times 10^{-5})$ in increasing thickness order. Right panel: time evolution
in models with $m_{\star}=2\times10^{-4}$ and for  $q=(5\times 10^{-3},10^{-2},5\times 10^{-2},10^{-1})$ in increasing thickness order of the black solid lines. {Dashed red curves show the theoretical prediction of Equation $(\ref{avinte})$ with
$\ln \left( 1/10^{-1}\right)$ while for the dashed blue curves we use $\ln \left( 1/5\times 10^{-3}\right)$}. Blue and purple thickmarks give the values of
 $a_{\rm crit}$ and $a_{\rm h}$ respectively. This figure shows that (i) the evolution above $a_{\rm crit}$ is not affected by the mass of the field particles, (ii) above $a_{\rm crit}$ dynamical friction leads towards higher eccentricities,
and (iii) below $a_{\rm crit}$ the eccentricity remains approximately constant with time.}\label{N-body1}
\end{center}
\end{figure*}

\subsection{Initial setup and numerical method}
We generate equilibrium $N$-body models of stars around a SMBH
adopting the truncated  mass model
\begin{equation}
\label{den2}
\rho(r)=\rho_0 \left(  \frac{r}{r_{\rm infl}} \right)^{-\gamma} 
\exp(-r/r_{\rm tr})\ .
\end{equation}
Monte-Carlo initial positions and velocities were assigned to the
$N$-body particles  by numerically  generating the 
distribution function corresponding to the isotropic
equilibrium model of Equation\ (\ref{den2}), while at the same time taking into account
the gravitational potential due to the central SMBH. A massive particle was placed in these models with mass $m\gg m_{\star}$, with corresponding mass ratio $q\ll 1$ at an initial galactocentric distance  $r\lesssim r_{\rm tr}$.

The initial conditions were evolved forward in time using the direct-summation code {\sc $\phi$GRAPE} \citep{2007NewA...12..357H}. 
This code uses a fourth-order Hermite integrator with a predictor-corrector scheme 
and hierarchical time steps.
The performance and accuracy of the code depend both on the time-step parameter $\eta$ and 
on the  smoothing length $\epsilon$. We set $\eta=0.01$ and $\epsilon=5\times 10^{-4}$.
With these choices, energy conservation was typically of order $0.1\%$ over the entire length of the integration.
The calculations were carried out in serial mode using graphics processing units combined with the {\sc sapporo} library \citep{2009NewA...14..630G,2015ComAC...2....8B}.

\subsection{Results}
Figure \ref{N-body0} shows the evolution of the semi-major axis and eccentricity 
of the massive binary in $N$-body models
with  $\gamma=(0.6,1.5,2)$, 
two values of binary mass ratio $q=(2\times10^{-4},\ 5\times10^{-5})$,
and two different values for the mass of the field particles
$m_\star=(5\times10^{-6}, \ 1\times10^{-5})$.
In all the models shown in  Figure \ref{N-body0} we set $r_{\rm tr}=0.2$,
 and place the secondary massive body
at $r=r_{\rm tr}$ with a velocity half the circular value. With this choice of parameters, the initial 
values of semi-major axis and eccentricity are
$a_0\approx 0.1$ and $e_0\approx 0.7$.
% The results of the  $N$-body simulations 
%are compared to the predictions of equations $(\ref{avinta})$ and $(\ref{avinte})$.

The resulting eccentricity and semi-major axis evolution in the 
simulations shown in Figure \ref{N-body0} are
in good agreement with the theoretical predictions.
 In all cases, the semi-major axis 
evolves more rapidly  than  the eccentricity does.
Indeed, we find no   systematic change in the binary eccentricity
over the simulated timescale for any value of $\gamma$. Although 
this behavior seems largely consistent with the predicted evolution, 
we also observe  random-like variations of the orbital eccentricity.
Such fluctuations in $e$ are due to hard scattering off surrounding 
stars which cause the angular momentum of the  massive body to 
random walk with an amplitude that decreases with increasing 
 mass ratio $m/m_\star$ \citep[e.g.,][]{2007ApJ...656..879M}. 
 We note that in a real galaxy the mass ratio between the 
secondary massive body and field stars could be much larger than in our simulations,
so such an effect would be essentially absent.

\begin{figure}
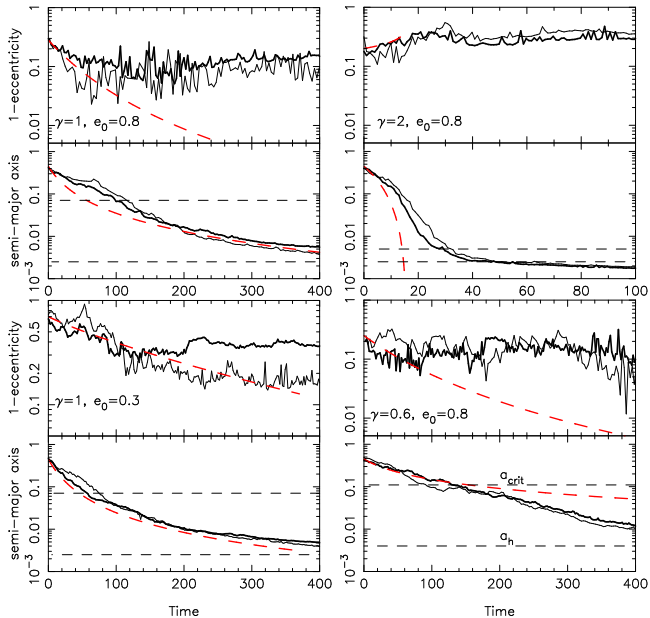

\begin{center}
   \includegraphics[angle=0,width=1.7in]{plot1.eps} 
   \includegraphics[angle=0,width=1.615in]{plot2.eps}
\includegraphics[angle=0,width=1.7in]{plot12.eps}
    \includegraphics[angle=0,width=1.615in]{plot06.eps} 
  \caption{Evolution of massive binaries with $q=10^{-2}$ in models with $m_{\star}=10^{-4}$ (thin lines)
and $2\times 10^{-4}$ (thick lines). Dashed red lines show the theoretical predictions based on Equation $(\ref{avinte})$. Dashed black lines depict the values of $a_{\rm crit}$ and $a_{\rm h}$.}\label{N-body2}
\end{center}
\end{figure}

 Figure \ref{N-body1} and \ref{N-body2}
show simulations where 
the orbit of the massive  body was followed 
 until $a$ had shrunk by a factor of $\approx100$ below its intial value.
In these additional simulations 
we increased the mass of
the secondary body and field particles, and set 
$r_{\rm tr}=1$.

In the simulations shown in Figure \ref{N-body1}
 $a_0\approx 1$ and $e_0 \approx 0.3$.
We see from the left panels that the orbital evolution of the binary at $a\gtrsim a_{\rm h}$
is essentially independent on $m_{\star}$, or equivalently on the number of
particles, $N$, used to represent  the galaxy. 
At later times, $t\gtrsim200$ or at $a\lesssim a_{\rm h}$, 
 the binary hardening rate becomes significantly longer and  
shows a clear $N$-dependence, in the sense of slower
 hardening for larger $N$.  In this phase the 
 hardening of the binary requires a repopulation of the depleted orbits through 
 collisional loss-cone repopulation which is an $N$-dependent process
  \citep[e.g.,][]{2002MNRAS.331..935Y}.

The right panels of Figure \ref{N-body1}
  show the evolution of the massive binary in a set of integrations with the same value of 
  $N$ but for different values of $q$.
{ From these plots we see that the dynamical friction timescale
of a massive binary above $a_{\rm crit}$  scales
approximately linearly with the mass of the secondary body (see also Figure \ref{N-body0}).}
This is also expected given that 
the dynamical friction acceleration decreases proportionally with $m$, 
if we neglect the logarithmic dependence of the frictional drag on $m$ through $\ln \Lambda$.
As before, the massive binary orbit is observed  to stall at  $\approx a_{\rm h}$.

Figure \ref{N-body1} and \ref{N-body2} show that the evolution of the binary eccentricity can be 
divided   in two distinct phases. At separations $a>a_{\rm crit}$ the eccentricity 
 of the binary changes steadily with time:
it increases for  $\gamma=(0.6,~1)$ 
and decreases for $\gamma=2$.
 In the second phase, at  $a<a_{\rm crit}$, the eccentricity established at early times tends to persist, remaining approximately 
constant as the orbit keeps shrinking.
 Because $a_{\rm crit}$ increases with $m$ [see Equation\ (\ref{acrit})], 
   lighter inspiraling bodies have higher eccentricities by the time their orbit has 
shrunk to $a\approx a_{\rm h}$.  This fact appears  evident in the top right panel of  Figure \ref{N-body1}
   where the general trend is clearly toward higher final eccentricities for smaller  $q$.

\subsection{Comparison to analytical predictions}

As can be seen from Figures \ref{N-body0}, \ref{N-body1} and \ref{N-body2} 
the evolution of $e$ and $a$ at $a>a_{\rm crit}$ in the $N$-body simulations show 
good agreement with the theoretical predictions. 
 For the cases we considered 
 the results of the $N$-body simulations confirm the expected
qualitative result  that for $\gamma\leq1$  the eccentricity will increase 
during the inspiral while for $\gamma \geq 2$  the eccentricity will decrease.
We conclude that the results of the  $N$-body simulations support
  the correctness of 
  the analytical treatment developed 
 in Section \ref{analytical} and consequently 
of Chandrasekhar's physical picture of dynamical friction.

Although the agreement with the theoretical prediction appears
fairly good a difference is also apparent: 
both $a$ and $e$ evolve more slowly in the  $N$-body simulations than predicted.
 This results in a small displacement towards the right of the corresponding analytical curves 
 shown in the figures. The discrepancy 
  is caused by  the fact that in our analytical treatment
 the contribution of the stars to the gravitational potential is ignored. Consequently,
the  massive body moves slower relative to the $N$-body which leads to 
an artificially  stronger frictional drag due to the $\propto v^{-2}$ dependence that appears 
in Equation (\ref{df}). 
This also explains why  for smaller initial separations
the agreement appears to improve --
for smaller  $a_0$ the contribution of the stars to the gravitational potential can be more safely
neglected given that the 
potential is more strongly dominated by the primary SMBH.
 
 We have shown that after the binary semi-major axis has decreased to $a_{\rm crit}$ 
 the eccentricity remains approximately constant in time. Therefore
 the value of $e$ at $a_{\rm crit}$ 
is also approximately the eccentricity the binary will have at the time it has decayed to $a_{\rm h}$. 
Below $a_{\rm h}$ the evolution becomes more uncertain.
Scattering experiments typically suggest a slow but steady growth of eccentricity \citep{1992MNRAS.259..115M}.

\begin{figure*}
  \centering
    \includegraphics[angle=0,width=6in] {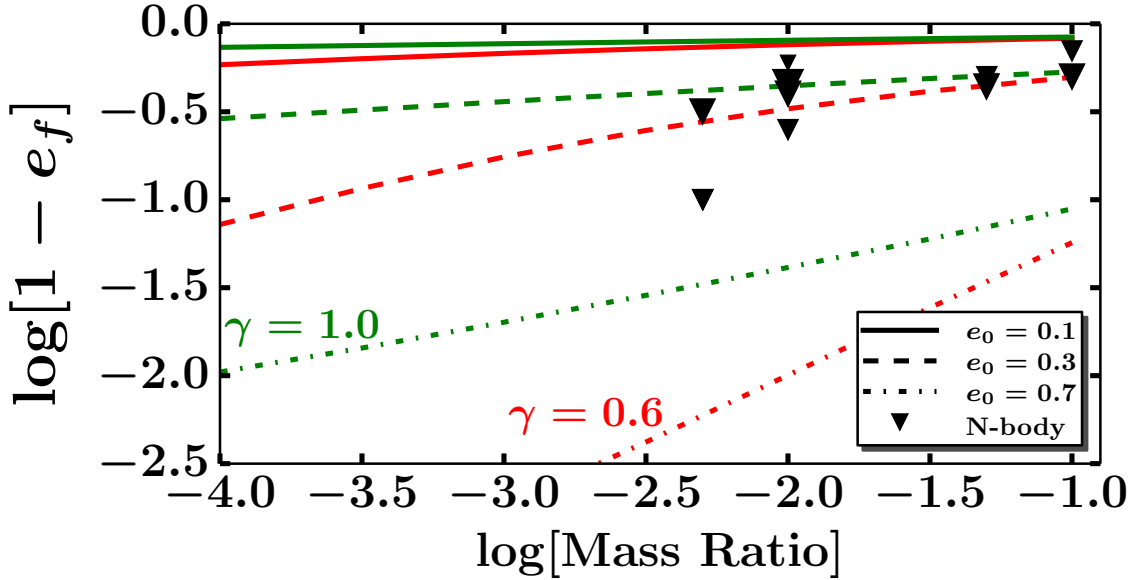}
      \caption{Eccentricity of the massive binary at the moment  the inspiraling body reaches $a_{\rm crit}$ as a function of the binary mass ratio. Black triangles represent the results from the $N$-body simulations shown in Figure \ref{N-body1} where $\gamma=1$, $e_0=0.3$ and $a_0=1$.
The values of $e$ from these simulations are averaged values at $a\leq a_{\rm crit}$.
Increasing marker size indicates larger
number of particles, i.e., lower $m_\star$.}
\label{finalecc}
\end{figure*}

 The predicted eccentricity of the massive binary at 
 $a_{\rm crit}$ is given in Figure \ref{finalecc} as a function of the binary mass-ratio,
 different initial values of the orbital eccentricity and for $\gamma=(0.6,~1)$. 
 In this plot we compare the theoretically predicted results with the results from the $N$-body simulations
 which are shown as solid triangles, and that were obtained as the average value of $e$ at radii $a<a_{\rm crit}$. 
 The agreement with the analytical predictions is again good.
 We see that for low mass-ratio binaries and a moderate initial eccentricity,  $e_0\gtrsim 0.3$, the 
 binary will reach $e\gtrsim 0.9$  by the time it has decayed to $a_{\rm crit}$.
The results of this analysis indicate that due to dynamical friction
the eccentricity of a comparatively low mass test body
moving in the center of a massive galaxy will be high during the time it spends inside the sphere of influence
of the central SMBH.
 
The eccentricity evolution itself is especially important in the case of SMBH binaries since the energy losses due to gravitational wave
 emissions depend strongly on $e$.
 How much the binary must shrinks by stellar-dynamical processes before the
 gravitational wave emission takes over
is very sensitive to the eccentricity of the binary. In what follows we apply the
 formalism developed in Section \ref{analytical} and confirmed with  $N$-body simulations
  to describe the evolution of SMBH binaries in early-type galaxies.

\section{Formation of SMBH binaries}\label{formation}
In what follows, we discuss the implications of our results
in relation to the dynamics
of the SMBH binaries that are believed to form during the merger of galaxies.

The evolution of a SMBH binary can be divided
into three phases:
(i)  the large scale orbital decay of the satellite galaxy from a distance of 
 order the primary galaxy half-mass radius, or effective radius. This phase ends when
the separation  between the two SMBHs reaches the 
primary SMBH influence radius, $\sim r_{\rm infl}$;
(ii) the inspiral of the satellite galaxy SMBH within the sphere of
influence of the primary SMBH. In this phase the motion of the secondary SMBH 
 is approximately Keplerian and the evolution of its orbit
can be described by Equations \ (\ref{da})-(\ref{dsigma}); (iii) 
when the binary's binding energy reaches $\sim M_\bullet\sigma^2$
the  two SMBHs form a ``hard binary'' at the center of the merger
product. At this stage, stars that intersect the SMBH binary are ejected from the 
system. We anticipate here that the evolution at this stage is likely to be efficient, leading to the
 hardening of the binary  and to its final coalescence on a timescale $\lesssim 1 \rm Gyr$.
 
In this section we give expressions to calculate the characteristic timescales associated with 
each of the three SMBH binary evolutionary stages, and discuss the connection between dynamical friction 
 and the formation of stalled SMBH satellites in luminous galaxies.

\subsection{Large scale orbital decay}

Modeling the main galaxy as a singular isothermal sphere, 
the  timescale for a satellite SMBH of mass $m$
to decay towards the center is \citep{1987gady.book.....B}:
\begin{eqnarray}\label{bare}
T^{\rm bare}_{\rm \star}={17}\frac{6.6}{\ln\Lambda}\ \left(R_e\over 10\rm kpc\right)^2\left(\sigma\over 300\rm km\ s^{-1}\right) \nonumber \\
\left(10^8M_\odot \over m\right)\ \rm Gyr
\end{eqnarray}
where $R_e$ is the effective radius of the main galaxy.

Some previous work made use of Equation\ (\ref{bare}) to 
describe the formation and evolution of SMBH binaries during galaxy mergers 
\citep[e.g.,][]{2014ApJ...789..156M}.
However, Equation\ (\ref{bare})
can lead to a significant overestimate of the real dynamical friction timescale and
 to erroneous conclusions about  the survival timescale of SMBH pairs
at the scale of $R_e$.
Equation\ (\ref{bare}) neglects the fact that the satellite  SMBH is brought in
during the course of a galaxy merger and it will retain some of its host galaxy's stars
until late in the merger process.
Considering the stellar mass bound to the inspiraling SMBH leads 
to a significantly shorter timescale for the formation of a bound pair.

By assuming a strict proportionality between the mass of the 
 stellar bulge of the satellite galaxy, $M_s$, and the mass of its central SMBH
$
M_s=10^{3}m 
$
\citep{2001MNRAS.320L..30M},
and replacing  $m$ with $M_s$ in Equation \ (\ref{bare}) gives
\begin{eqnarray}\label{gx1}
T^{\rm gx}_{\rm \star,1}={0.06}\frac{2}{\ln\Lambda'}\ \left(R_e\over 10\rm kpc\right)^2\left(\sigma\over 300\rm km\ s^{-1}\right) \nonumber \\
\left(10^{8}M_\odot \over m\right)\ \rm Gyr\ ,
\end{eqnarray}
where the argument of the Coulomb logarithm is taken to be 
$\Lambda' = 2^{3/2}\sigma/\sigma_s$ with $\sigma_{s}$ the stellar velocity dispersion of the secondary galaxy.

Equation (\ref{gx1}) does not consider that the satellite galaxy can be stripped of its stars due to the
strong tidal field of the primary galaxy.
Following \citet{1987gady.book.....B} we assume that  the satellite galaxy can also be modeled as a singular
isothermal sphere, so that its mass is related to its velocity dispersion
through the relation:
\begin{equation}\label{msec}
M_{\rm s} (r)\approx \frac{1}{2G}\alpha^2\sigma_s^2r_t \ .
\end{equation}
Setting the Hill radius 
$r_t=\left( G M_m r^2/4 \sigma^2\right)^{1/3}$  with 
$M_m$ the mass of the main galaxy and
 $\alpha=2$ as appropriate for a sharp truncation, the dynamical friction 
timescale becomes:
\begin{eqnarray}\label{totdf}
T^{\rm gx}_{ \star,2}={0.15}\frac{2}{\ln\Lambda'}\ 
\left(R_e\over 10\rm kpc\right)\left(\sigma\over 300\rm km\ s^{-1}\right)^2 \nonumber \\
\left({100\rm km\ s^{-1}} \over \sigma_s \right)^3\ \rm Gyr.
\end{eqnarray}

A good approximation to the timescale for a secondary SMBH 
to decay from $R_e$ to  the influence radius of the primary SMBH,
is
\begin{equation}\label{tdf1}
T_{\rm \star}=\max(T^{\rm gx}_{\rm \star,1},T^{\rm gx}_{\rm \star,2})\ .
\end{equation}

\subsection{Dynamical friction of a bound pair}
As the secondary SMBH enters the sphere of influence of the primary,
the SMBHs are bound to each other and the formulae given above, which
are only strictly valid for a Maxwellian distribution of velocities and
a self-gravitating cluster, can no longer
apply. Nevertheless, most work in the past \citep[e.g.,][]{2016arXiv160601900K}
has neglected such complication and 
applied Equation\ (\ref{bare}) until $a_{h}$.

Although such simplification is reasonable for major mergers,
we find that for $q\lesssim 0.1$ 
it necessarily leads to an erroneous evaluation of the binary evolution
 timescale as well as the evolution of its orbital eccentricity.

Using Equation  (\ref{da}) it can be shown  that 
the characteristic dynamical friction
 timescale to decay from the primary SMBH influence radius, $r_{\rm infl}$,
 to a much shorter separation $r= \chi r_{\rm infl}$,  is:
 \begin{eqnarray}\label{bhdf}
T^{\rm bare}_{\bullet}=1.5\times 10^7
{\left[\ln \Lambda \alpha + \beta+\delta\right]^{-1}\over (3/2-\gamma)
(3-\gamma)} \left(\chi^{\gamma-3/2}-1\right)\\
\left(\frac{M_\bullet}{3\times 10^9 M_\odot}\right)^{1/2}
\left(\frac{m}{10^8 M_\odot}\right)^{-1}
\left(\frac{r_{\rm infl}}{300\rm pc}\right)^{3/2}  \nonumber
\rm yr \ .
 \end{eqnarray}
The coefficients $\alpha$, $\beta$ and $\delta$ can be easily  computed
from Equation (\ref{int3}), (\ref{int4}) and (\ref{int5}),  assuming a circular orbit,
i.e., setting  $\xi=1$ in these equations as justified by the fact that the 
orbital decay timescale is 
not significantly affected by the binary orbital eccentricity.

 Equation\ (\ref{bhdf}) does not 
consider that part of the host galaxy can remain bound to the secondary SMBH even inside 
$r_{\rm infl}$.
Assuming as before that the satellite galaxy can be modeled as a singular
isothermal sphere, then the 
 timescale to decay from $r_{\rm infl}$,
 to a smaller radius $r=\chi r_{\rm infl}$, can be obtained 
 by replacing $m$ with $M_s(r)$, 
where now $r_t\approx (M_s/2M_\bullet)^{1/3}r$.
In this case, Equation (\ref{da}) leads to
  \begin{eqnarray}
  T^{\rm gx}_{\bullet}=1.2\times 10^7 {\left[\ln \Lambda' \alpha + \beta+\delta\right]^{-1} 
 \over (3-\gamma)^2} \left(\chi^{\gamma-3}-1\right)\\
 \left(\frac{M_\bullet}{3\times 10^9 M_\odot}\right)
  \left({100\rm km\ s^{-1}} \over \sigma_s \right)^3\ \nonumber\ 
 \rm yr .
\end{eqnarray}
The decay timescale from  the influence radius of the primary SMBH 
is then 
\begin{equation}\label{tdf-rinf}
T_{\rm \bullet}=\min(T^{\rm bare}_{\bullet},T^{\rm gx}_{\bullet})\ .
\end{equation}

Note that Equation (\ref{tdf-rinf}) is strictly valid  only 
until the secondary hole reaches $a_{\rm crit}$ (see Equation (\ref{acrit})).
Below this radius, the central cusp starts to be significantly modified by the 
secondary SMBH \citep{2006MNRAS.372..174B} and time-dependent galaxy models  are required in order to 
describe the detail evolution of the
binary orbit \citep{2012ApJ...745...83A}. 
Here we ignore such complication and set $\chi= a_{\rm crit}/r_{\rm infl}$. Our choice is clearly  conservative.

\begin{figure}
\begin{center}
 \includegraphics[angle=0,width=2.5in]{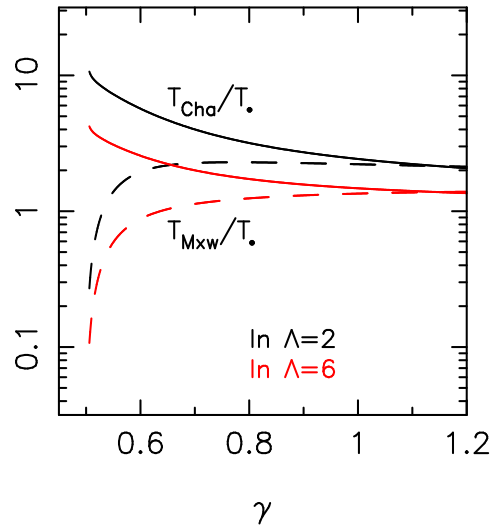}
    \caption{The solid lines give the
dynamical friction
timescale to reach $a_{\rm crit}$, $T_{\rm Cha}$,  derived from Equation\ (\ref{bhdf}) by setting $\beta=\delta=0$ and
divided by our more accurate estimate $T_\bullet$.
The dashed lines are the dynamical friction timescale, $T_{\rm Mxw}$,
    computed using the Maxwellian  approximation
     and divided by $T_{\bullet}$.    This calculation quantifies the error one would make by employing 
 the standard Chandrasekhar's formula  in which the contribution 
of the fast stars is neglected, or by assuming that the velocities of the field stars follow a Maxwellian distribution.
     } 
\label{compare}
\end{center}
\end{figure}

{
In order to quantify the error one would make by employing 
 the standard Chandrasekhar's formula, we plot in   Figure \ref{compare}
 the dynamical friction
timescale  from Equation\ (\ref{bhdf}) with $\beta=\delta=0$  
divided by $T_\bullet$.
For  
$\gamma\lesssim  1$
the standard Chandrasekhar's theory can lead to  significant 
deviations  from our more accurate 
formulation. Neglecting the fast moving stars' contribution  leads
to an overestimate of the dynamical friction timescale that can be
longer than $T_\bullet$ by about one order of magnitude for  $\gamma\approx 0.5$.
In Figure \ref{compare} we also compare our estimate to that obtained by
 assuming that the velocity distribution of the
field stars follows a Maxwellian distribution. 
This latter approximation underpredicts the decay timescale  by about one order of magnitude for  $\gamma\approx 0.5$.
For $\ln \Lambda\gtrsim 6$ and/or  $\gamma\gtrsim 1$ both approximations
give a good estimate of the decay timescale, as 
 the contribution of the fast stars to the drag force
is smaller in this case.
}

% Finally,  we find that 
%$T^{\rm bare}_{\bullet}<T^{\rm gx}_{\bullet}$  in all cases we considered,
%as expected if most of the host galaxy mass is removed before 
%reaching $a_{\rm h}$.
%Hence, in what follows we simply compute the decay  timescale  
%from $r_{\rm infl}$ to $a_{\rm h}$ as $T_\bullet=T^{\rm bare}_\bullet$.

\begin{figure*}
\begin{center}
 \includegraphics[angle=270,width=6.in]{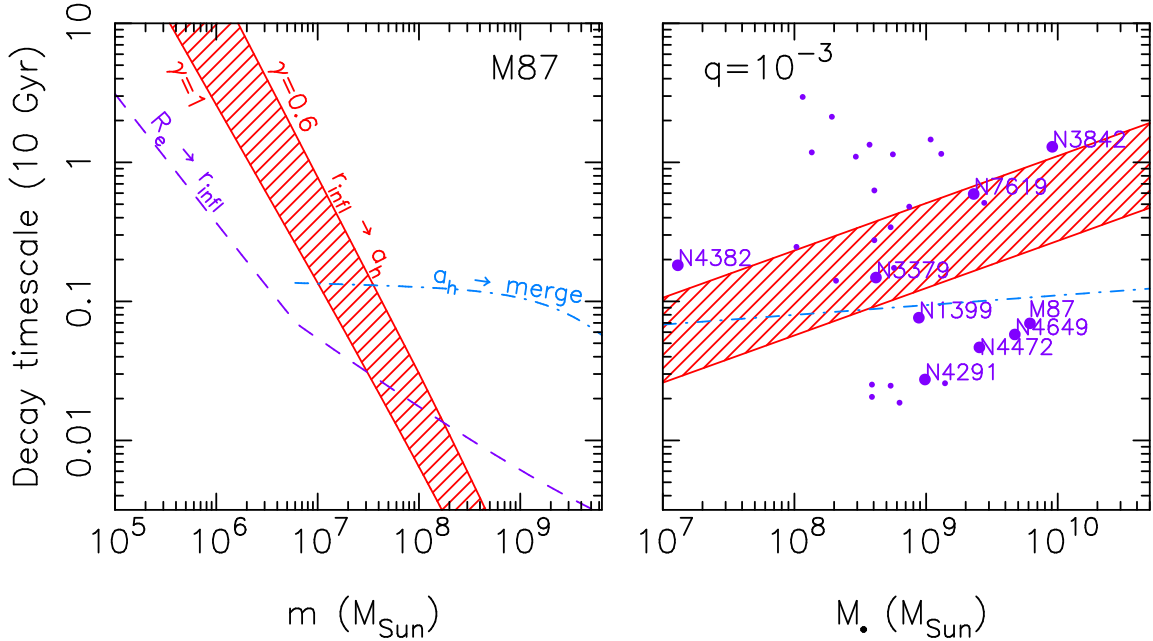}    
  \caption{Left panel:
  orbital decay timescale of  a secondary SMBH  as a function of its mass, $m$, in the nucleus of M87.
  Purple-dashed line is from Equation (\ref{tdf1}) and gives the decay timescale from the effective radius of the galaxy, $R_e$, to the sphere  of
  influence of the primary SMBH. Red-hatched region gives the timescale  to decay from $r_{\rm infl}$ to $a_{\rm crit}$. This
  timescale was computed via Equation (\ref{tdf-rinf}) setting $\gamma=1$ (left-red curve) and $\gamma=0.6$ (right-red curve) as representative values for the
density profile slope of the inner core of M87. Blue-dot-dashed line is the hardening timescale to decay from
    $a_h$ to coalescence computed using Equation\ (\ref{vam15}),
     assuming triaxial geometry and
a moderate eccentricity $e=0.3$.
For $m\lesssim 10^8\ M_\odot$ we find
$T_\bullet > T_{\rm h}> T_\star$.
Right panel:
orbital decay timescales as a function of $M_\bullet$ for
 $q=10^{-3}$.
Purple points  give the dynamical friction timescale
 $T_\star$
for the core-S\'{e}rsic galaxies in \citet{2015ApJ...798...55D}.
Larger points are systems with a direct SMBH mass measurement \citep{2013ARA&A..51..511K}.
As before, red-hatched region gives the timescale
  to decay from $r_{\rm infl}$ to $a_{\rm crit}$ for $\gamma=0.6$
  and $\gamma=1$.  Blue-dot-dashed line is $T_{\rm h}$.}
\label{binary-decay}
\end{center}
\end{figure*}

\subsection{Hardening}

The phase of binary evolution 
determined by dynamical friction 
comes to an end when the binary's  semi-major axis reaches
$\sim a_{\rm h}$.
Gravitational encounters will continue supplying stars to the binary at a rate
that depends on the host galaxy morphology.

{ 

Dry mergers  between  luminous galaxies result in triaxial remnants, 
which  leads to an efficient hardening of the binary \citep{2011ApJ...732...89K,2014ApJ...785..163V}.
After a major merger   a galaxy is likely to  retain some degree of triaxiality or
flattening during its subsequent evolution.
The timescale to decay from 
$a_{\rm h}$ to coalescence is \citep{2015ApJ...810...49V}
\begin{eqnarray}\label{vam15}
T_{\rm h}&\approx& 1.2\times 10^9  \left(r_{\rm infl}\over 300\rm pc \right)^{10+4 \psi\over 5+ \psi}
\left(M_\bullet+m\over 3\times 10^9 M_\odot\right)^{-5-3 \psi \over 5+ \psi} \\
&& \phi^{-4\over 5+ \psi}\left(4q\over (1+q)^2\right)^{3 \psi-1\over 5+ \psi}p(e)
\rm\ yr
\end{eqnarray}
with 
\begin{eqnarray}
p(e)=(1-e^2)\left[k+(1-k)(1-e^2)^4\right], \\
k=0.6+0.1\log\left([M_\bullet+m]/3\times10^9 M\odot \right) . \nonumber
\end{eqnarray}
The parameters $\phi$ and $ \psi$ parameterize the evolving hardening rate 
with values estimated from Monte Carlo simulations.
 In what follows we adopt  $\phi=0.4$ and $ \psi=0.3$ %$\phi=0.4$ and $\delta=0.3$ 
which are the values derived by \citet{2015ApJ...810...49V} for triaxial galaxies.

We describe how much
the binary must shrink by stellar-dynamical processes before
the GW emission takes over using the ratio \citep{2015ApJ...810...49V}:
\begin{eqnarray}\label{gwd}
\nonumber {a_{\rm h}\over a_{\rm GW}}\approx 55\times 
\left( {r_{\rm infl}\over {30\rm pc}}\right)^{{5 \over 10}}  
\left(M_\bullet+m \over 10^8M_\odot\right)^{-{5\over10}}
\\
\times f(e)^{1\over 5} \left( 4q\over (1+q)^2 \right)^{4\over 5}
%\left({M_{\bullet}+m \over 10^8 M_{\odot}} \right)^{-{5\over10+2\delta}}
%\nonumber {a_{\rm h}\over a_{\rm GW}}\approx 160\times 
%\left( {r_{\rm infl}\over {30\rm pc}}\right)^{{5 \over 10+2 \delta}}  
%\left(M_\bullet+m \over 10^8M_\odot\right)^{-{5\over10+2\delta}}
%\\
%\times [\phi f(e)]^{1\over 5+\delta} \left( 4q\over (1+q)^2 \right)^{4\over 5+\delta}0.4^\delta
%\left({M_{\bullet}+m \over 10^8 M_{\odot}} \right)^{-{5\over10+2\delta}}
\end{eqnarray}
with 
\begin{equation}
f(e)={{(1-e)^{-{7\over2}}}\over {1+{73\over 24}e^2+{37\over96}e^4}}
\end{equation}
and $a_{\rm GW}$ the separation at which gravitational wave radiation takes over.
From Equation\ (\ref{gwd}) one finds that $a_{\rm h}<a_{\rm GW}$ for $q\approx 10^{-3}$ and moderate eccentricities.
Binaries with mass ratio lower than this value never enter the hardening phase.
The quoted
approximation formula therefore breaks down in this case because these systems are never in a
properly stellar-dynamical hardening regime.
These binaries transit directly from the dynamical friction phase to the phase where
gravitational wave  radiation leads to their rapid coalescence.  
Accordingly, in what follows we set $T_{\rm h}=0$ for $q\lesssim 10^{-3}$.

 Equation (\ref{vam15}) implies that stellar-dynamical interactions are  able to drive the binary to 
coalescence on a timescale $\leq 1$Gyr in any triaxial galaxy, and that the coalescence timescale weakly depends on the mass of the binary, and its mass ratio.
Coalescence times also depend quite significantly on the binary eccentricity
 falling in the range from a few Gyr for almost circular binaries, to  $\approx 10^8$ yr for very eccentric ones.

We note that Equation\ (\ref{vam15}) was derived for
major mergers, $q\gtrsim 0.1$.  However, additional simulations 
performed using the Monte-Carlo code RAGA \citep{2015MNRAS.446.3150V} showed that 
 Equation\ (\ref{vam15}) works remarkably well for  all binaries 
with $q\gtrsim 10^{-2}$. At  $q\approx 10^{-3}$ 
 Equation\ (\ref{vam15}) 
 starts to break down, but even for $q=10^{-3}$ it
 overestimates by only a factor of $\approx 3$ 
 the binary's coalescence timescale (E. Vasiliev; private communication).
 Hence, unless otherwise specified, in what follows we adopt Equation\ (\ref{vam15})
in the full range of mass ratios $10^{-3}\leq q\leq1$.
The fact that $T_h$ is only an order of magnitude estimate for 
  the lowest mass ratio binaries  we considered does not affect 
our calculations below, given that the total lifetime of these binaries is largely 
dominated by their dynamical friction timescale.
 }

Finally, we note that the assumption of triaxiality in the field-star distribution made in Equation\ (\ref{vam15})
is obviously inconsistent with the spherical density model 
used to compute the dynamical friction timescales above.
However,  
galaxy mergers produce remnants 
with deviations from isotropy that are small both in velocity and configuration space  \citep{2015ApJ...810...49V}.
Hence, we might expect our estimates of the dynamical friction timescales to give a reasonable approximation.
\begin{figure*}
\begin{center}
 \includegraphics[angle=0,width=2.3in]{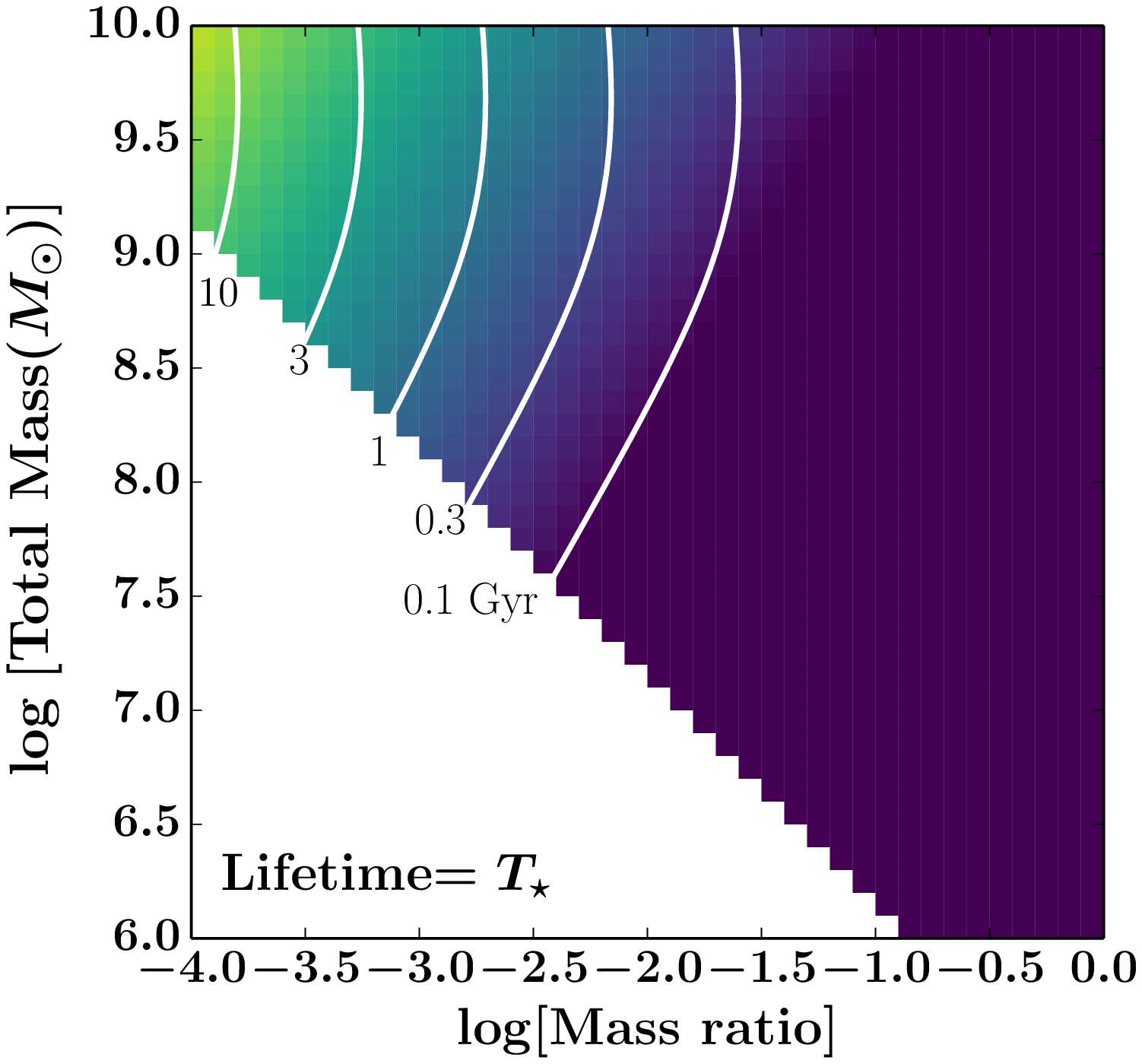}
   \includegraphics[angle=0,width=2.1in]{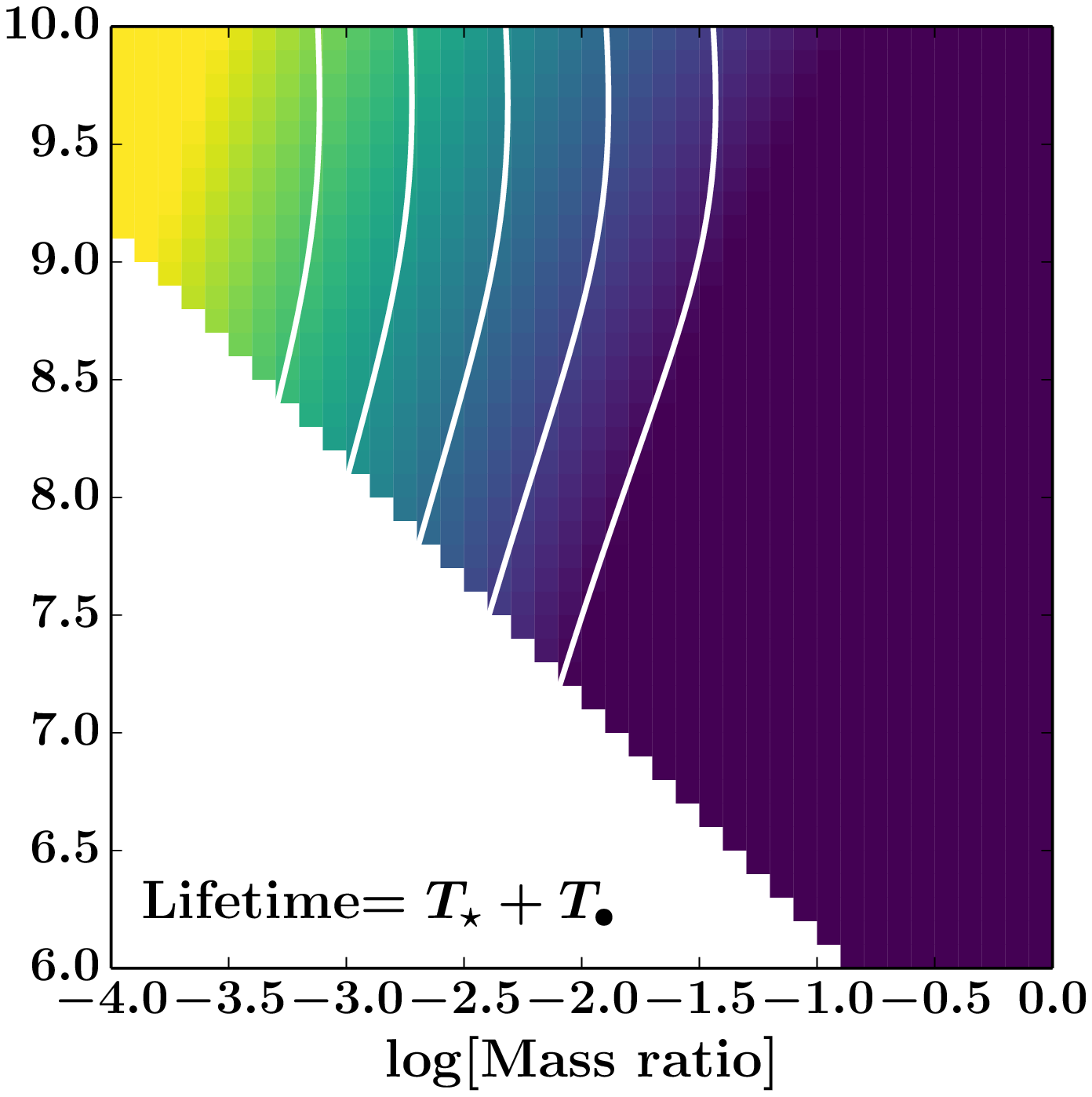}
   \includegraphics[angle=0,width=2.5in,height=2.1in]{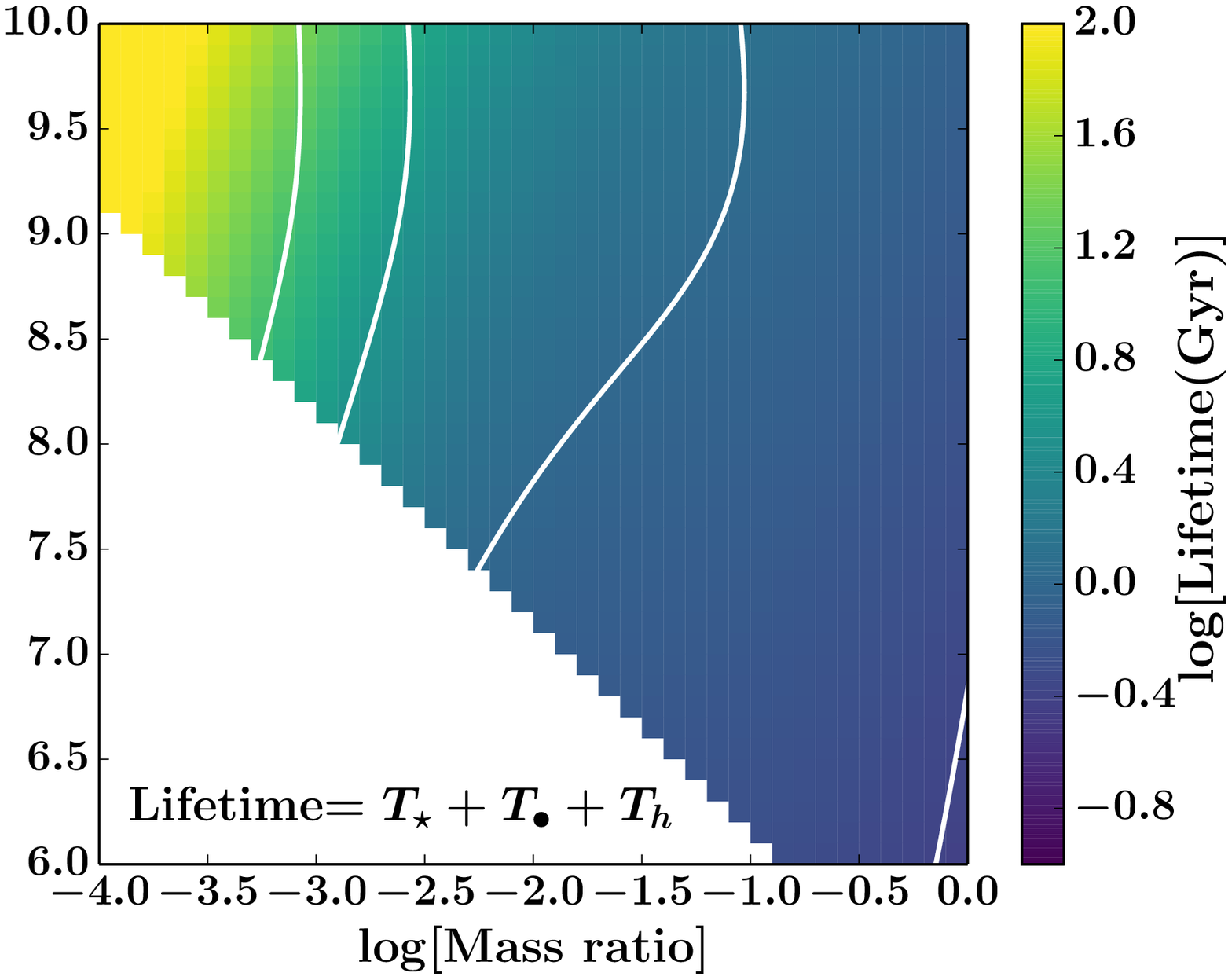}
  \caption{Lifetime of a SMBH binary (colorbar) as a function of its total mass and mass ratio. We set $\gamma=0.6$. Left panel shows the time for a SMBH binary  to decay from $R_{\rm eff}$ to $r_{\rm inf}$, central panel to decay from $R_{\rm eff}$ to $a_{\rm crit}$, and right panel from $R_{\rm eff}$ until coalescence. The lifetime of a SMBH binary to the left of the solid white lines is above $10,3,1,0.3$ and $0.1$ Gyr respectively. The lifetime of SMBH binaries with a low mass ratio becomes significantly
longer when we include in the calculation the timescale for dynamical friction inside $r_{\rm inf}$ which is often neglected in the literature.}
\label{lifetime}
\end{center}
\end{figure*}

\subsection{SMBH binary formation in early-type galaxies}

Here we estimate the timescales associated with the three stages of binary evolution defined above
for real galaxies and consider the possibility of 
``stalled'' mergers in these systems, where
a smaller satellite SMBH 
resides in the outer regions of the
main host galaxy. 
We focus on massive early type galaxies because
 (i) galaxy mergers are thought to play a crucial role in the late growth 
 of these systems and,
  (ii) these galaxies are often observed to have 
extended  density profile cores within 
the sphere of influence of the SMBH, which implies a long dynamical friction timescale.
 For these reasons, luminous early-type galaxies are likely hosts of stalled satellite SMBHs.
{ On the other hand  our results imply that in low mass ellipticals, SMBH binaries
have a short timescale, i.e., shorter inspiral times, possibly
 in the decreasing or near constant eccentricity regime.}

We start by considering the case of a widely studied
 massive galaxy such
as M87, for which $M_\bullet\approx 6.15\times 10^9M_\odot$,
 $\sigma \approx 330 \rm km\ s^{-1}$ \citep{2013ARA&A..51..511K}, $R_e\approx 9\rm kpc$ \citep{1995AJ....110.2622L} and 
 $r_{\rm infl} \approx 300\rm pc$ \citep{2003ApJ...589L..21M}.
Given these structural parameters, the calculation of the decay timescales
can be simply performed
by assuming that the secondary galaxy velocity dispersion is related to 
the mass of its central SMBH through the $M_{\bullet}-\sigma$ relation defined below in Equation (\ref{sigmaz}) setting $z=0$ \citep{FM00,2000ApJ...539L..13G}
%\begin{equation}\label{msigma}
%{M\over 10^8 M_\odot}\approx 1.66\left(\sigma\over 200\rm km\ s^{-1}\right)^5 \ .
%\end{equation}

An additional parameter which needs to be defined is $\gamma$, the slope of the deprojected 
spatial density profile of stars within $r_{\rm infl}$.
We  adopt  the two representative values $\gamma =0.6$
and $1$,  as motivated by the facts that the M87 density profile 
 is  observed to be nearly flat inside $r_{\rm infl}$
and values $\gamma \leq 0.5$ are excluded by our assumption 
of kinematical isotropy.

The left panel of  Figure \ref{binary-decay} gives the orbital decay timescale
as a function of the satellite SMBH mass inside M87 at the different stages
of the evolution of the massive binary.
The dynamical friction timescale to decay from the 
effective radius to the SMBH influence radius, $T_\star$, is shorter than
$\approx 10^9\rm yr$  for $m\gtrsim 5\times 10^6$
implying that  SMBH binaries are unlikely  to stall near $R_e$.
The  hardening timescale to reach coalescence from the hard binary separation is
quite short  $\lesssim 10^9\rm yr$, even
for the moderate eccentricity we adopted $e=0.3$. 
The dynamical friction timescale to decay from $r_{\rm infl}$ 
to $a_{\rm h}$  can instead be
extremely long.
 For $m\lesssim 10^8 M_\odot$, $T_\bullet$
 becomes longer than either $T_{\star}$ or $T_{\rm h}$,
and it is longer than $10\rm Gyr$   for $m\lesssim 10^6 M_\odot$ ($m\lesssim 10^7 M_\odot$ )  when $\gamma =1$ ($\gamma =0.6$).

In the right panel of Figure \ref{binary-decay}
we show the timescale of orbital decay  $T_\star$
for  the  sample of 31 core-S\'{e}rsic early-type galaxies in
\citet{2015ApJ...798...55D} (purple points). These systems
are bright elliptical and lenticular galaxies with  extended density profile cores. 
\citet{2015ApJ...798...55D} give the measured
structural parameters of these galaxies, including $R_e$
and $\sigma$, which we used  to compute $T_\star$.
This timescale is plotted
 as a function of the primary
SMBH mass for $q=10^{-3}$.
Larger symbols are  systems for which a direct SMBH measurement is available in 
the literature \citep{2013ARA&A..51..511K}. 
The resulting $T_\star$ is $\lesssim 10^9\rm yr$.
The  hardening timescale $T_{\rm h}$ (dot-dashed line) 
also appears to be quite short $\lesssim 10^9\rm yr$, and, as expected, weakly dependent on 
the primary SMBH mass.
Finally, the hatched red region gives $T_\bullet$ which was computed through Equation (60) by
setting  \citep[e.g.,][]{2009ApJ...699.1690M}
\begin{equation}\label{inflfit}
r_{\rm infl}\approx 35 \left(  M_\bullet \over 10^8 M_\odot   \right)^{0.56} \rm pc .
\end{equation}
with $M_\bullet$ given by Equation\ (\ref{sigmaz}) at $z=0$.

The analysis shown in the right panel of  Figure \ref{binary-decay}
confirms and generalizes our statement
 that the dynamical friction timescale inside the influence radius of a SMBH
can become of the order the Hubble time in luminous spheroids. 
Not just in  the core of M87 but for most galaxies in the sample we considered 
$T_\bullet$ can be significantly 
larger  then either $T_\star$  or $T_{\rm h}$, and can become longer than the Hubble time
even for relatively high mass ratio binaries. {The overall trend indicating longer lifetimes for 
higher total mass galaxies results from the strong dependence of Equations (\ref{bare}), (\ref{gx1})
and (\ref{totdf}) on the galaxy effective-radius.}

In Figure \ref{lifetime} we investigate further the dependence of the lifetime of a SMBH binary on its total mass and mass ratio. 
The total lifetime of a SMBH binary is given by $t_{\rm L}=T_{\star}+T_{\bullet}+T_{\rm h}$.
We plot the lifetime of a massive black hole binary in the three 
different stages of the binary evolution as a function of the binary total mass and mass ratio. 
For the effective radius of the host galaxy we use
$
 R_{\rm e}=1.35  \left( {M_{\bullet}}/{10^{8} M_\odot}\right)^{0.73}{\rm kpc}\label{rrz}
$
where the mass dependence and normalization  were obtained from the observed 
mass-radius relation of  galaxies at redshift zero {(Figure 7 in \citet{2008MNRAS.389.1924F} for their sample of elliptical galaxies)},  and after expressing the galaxy mass 
as $M_{\rm m}= 10^{3}M_{\bullet}$, assuming the latter relation holds at all redshifts.

Based on Figure \ref{lifetime} the amount of time the secondary SMBH spends to  decay from the effective radius of the galaxy $R_{e}$ to the influence radius of the central black hole is short, typically less than $\sim 3$ Gyr. Adding the time the binary spends in the dynamical friction phase until $a_{\rm crit}$ leads to
significantly longer binary lifetimes. Figure \ref{lifetime} shows that including the dynamical friction timescale for high total mass and low mass ratio binaries expands the parameter space for long-lived binaries with lifetimes greater than $\sim 3$ Gyr, while there is a considerable amount of these binaries that have
 lifetimes greater than $\sim 10$ Gyr. This implies that binaries with high total mass and low mass ratio  are expected to live long in the evolutionary stage between the SMBH influence radius and hardening radius of the binary. Based on Figure \ref{finalecc} we also expect that these binaries  will have high eccentricities. The final phase in the evolution of a SMBH binary before coalescence is characterized by the hardening timescale which we add in the binary total lifetime in the right panel of Figure \ref{lifetime}. The hardening timescale contributes mostly to the  lifetime of high mass ratio binaries, being always shorter than $\approx 1\rm Gyr$.

\section{Stalled satellites in minor mergers}\label{stalledsection}

In section \ref{formation} we studied the lifetime of a SMBH binary. We found that SMBH binaries with  high total mass and low mass ratio   are long-lived. These binaries are the product of minor galaxy mergers where a smaller galaxy is accreted by a giant galaxy. If the lifetime of a massive binary exceeds the time passed from its formation redshift to the present time the binary orbit stalls and the secondary SMBH becomes a \emph{stalled satellite}. The number of stalled satellites over the Hubble time depends on the formation redshift of the binary and the rate at which galaxies in the relevant total mass and mass ratio range merge with each other.

The total galaxy merger rate is defined as the rate at which a galaxy with a primary SMBH of mass $M_{\bullet}$ experiences mergers with other galaxies
at a redshift $z$.  The merger rate therefore depends on how the galaxy properties evolve with redshift.
In order to model the evolution of mass, effective radius and  velocity dispersion of a galaxy
we follow below the redshift-dependent fitting formulae of \citet{2012MNRAS.422.1714N} {for typical massive early-type galaxies}
\begin{align}
\mathcal{M}(z)&= M_{\bullet}(1+z)^{-0.6}\: M_{\sun}\label{mz}\\
\mathcal{R}_{e}(z)&=1.35 \left( \frac{M_{\bullet}}{10^{8}}\right)^{0.73}(1+z)^{-0.71}\:\rm kpc\label{rz}\\
\mathcal{\sigma}(z)&=180 \left( \frac{M_{\bullet}}{10^{8}}\right)^{0.2}(1+z)^{0.056} \:\rm km/s \label{sigmaz}
\end{align}
where now $M_{\bullet}$ indicates the central SMBH mass at redshift $z=0$.
 In what follows unless otherwise specified we will use Equations (\ref{mz})-(\ref{sigmaz}) as our reference model.

The differential merger rate of galaxies can be derived based on the differential fraction
of galaxies with central SMBH mass $M_{\bullet}$ at redshift $z$ that are paired with a
secondary galaxy having a mass ratio in the range $q$, $q + dq$, and the
merger timescale for a galaxy pair with a given $M_{\bullet}$ and $q$ at a given $z$ \citep[e.g.,][]{2008ApJ...685..235P,2008MNRAS.391.1489K,2009A&A...498..379D,2009ApJ...697.1369B,2012ApJ...747...85X,2012A&A...548A...7L}. In this paper we use the pair fraction derived in \citet{2012ApJ...747...85X}. The differential galaxy merger rate per unit mass ratio is then given by \citep{2012ApJ...747...85X,2013MNRAS.433L...1S,2016arXiv160607484R}
\begin{align}
\frac{dN_{\rm m}(z,q)}{dq\:dt}&=0.044 \frac{1}{q}\left(\frac{\mathcal{M}(z)}{10^{7.7}M_{\sun}}\right)^{0.3}
\frac{(1+z)^{2.2}}{1+z/8}\rm Gyr^{-1}\label{rate}
\end{align}
where $N_{\rm m}$ is the the number of mergers. {In the work of \citet{2012ApJ...747...85X}, Equation (\ref{rate}) refers to mergers in the data sample with maximum merger mass ratio $M_{\bullet}/m=2.5$. For the observed pair distribution as a function of the mass ratio $q$ a good proxy is  $\propto 1/q$ \citep[e.g.,][]{2013MNRAS.433L...1S}.  Cosmological simulations in \citet{2015MNRAS.449...49R} suggest that $\propto 1/q$ is also a good proxy for all mass ratios in the range $10^{-4}<q<1$. More specifically,  in \citet{2015MNRAS.449...49R}  $dN_{m}/dq_{h} \propto q_{h}^{-1.7}$, where $q_{h}$ refers to the dark matter  halo mass ratio. Following the work of \citet{2003MNRAS.341L..44B} and assuming a SMBH-halo mass relation  $M_{\bullet}\propto M_{h}^{1.3}$ leads to $dN_{m}/dq \propto q^{-1.5}$ (i.e., between 
the linear scaling adopted here and a quadratic scaling).}

 % \citet{2013MNRAS.433L...1S}  following the work of \citep[e.g.,][]{2009A&A...498..379D,2009ApJ...697.1369B,2012ApJ...747...85X,2012A&A...548A...7L}
  %Observations suggest that elliptical galaxies form their stars quickly and early $z>2.5$ and subsequently quench their star formation \citep[e.g.,][]{2008ASPC..399..260D}. They show early merging activity and then at $z<1.5$ not much growth by major merging is happening any more (at most a factor of 2 in stellar mass,1 effective merger per object in the redshift range $0<z<1.5$). At more moderate masses $7<\log(M_{\bullet})<8$ the evolution of the number density due to mergers point toward these galaxies being preferably destroyed at early times while at later times the change in their number density turns positive. At lower redshift the number density of progressively lower mass galaxies starts to increase due to merging. This means that we expect minor mergers and the formation of stalled supermassive black hole binaries to happen in the early redshift range $0<z<1$. For this reason we choose here for the maximum redshift $z_{\rm max}=1$,
%with the time passed from this maximum formation redshift to redshift zero being $t_{\rm H}(z_{\rm f}=1) \approx 10\: \rm Gyr$.
\begin{figure}
  \centering
    \includegraphics[width=.4\textwidth]{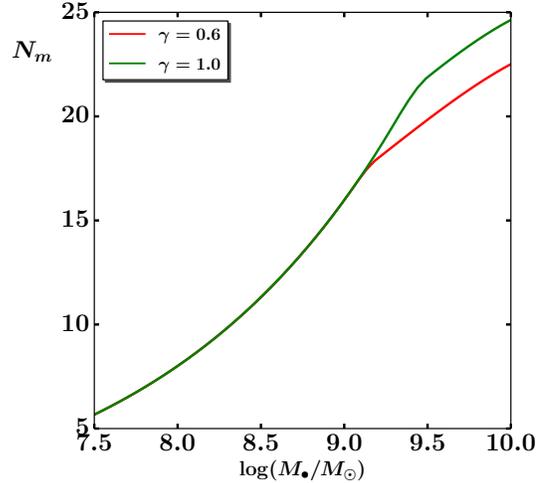}
      \caption{Average number of infalling satellites for $0<z<$1 and as a function of the host galaxy 
      SMBH mass for two different stellar cusp density profiles. }
\label{mergers}
\end{figure}
We can calculate the expected average number of stalled satellite SMBHs, $N_{\rm s}$, for a galaxy with a central SMBH with mass $M_{\bullet}$, by integrating the differential merger rate (\ref{rate}) over the relevant range of redshift $0<z<z_{\rm max}$ and mass ratio $q_{\rm min}<q<q_{\rm crit}$:
% \begin{align}
%N_{\rm s} (M_{\bullet})= \int_{0}^{z_{\rm max}}\int_{q_{\rm min}}^{q_{\rm crit}}  \frac{dN_{\rm m}}{dq\:dt}\frac{dt}{dz}dzdq
%\label{stalled}
%\end{align}
\begin{align}
N_{\rm s} (M_{\bullet})= \int_{0}^{z_{\rm max}}\int_{q_{\rm min}}^{q_{\rm crit}}  \frac{dN_{\rm m}(z,q)}{dq\:dt}\frac{dt}{dz}dzdq
\label{stalled}
\end{align}
where $q_{\rm crit}$ is  the critical value of the mass ratio below which the lifetime of the massive binary exceeds  the time passed from its formation until today. This was computed as the solution to the equation
\begin{equation}
t_{\rm L}(q_{\rm crit},z,\gamma)=10\rm \:Gyr\label{qcrit}\ .
\end{equation}
\begin{figure*}
\begin{center}
 \includegraphics[angle=0,width=3.3in]{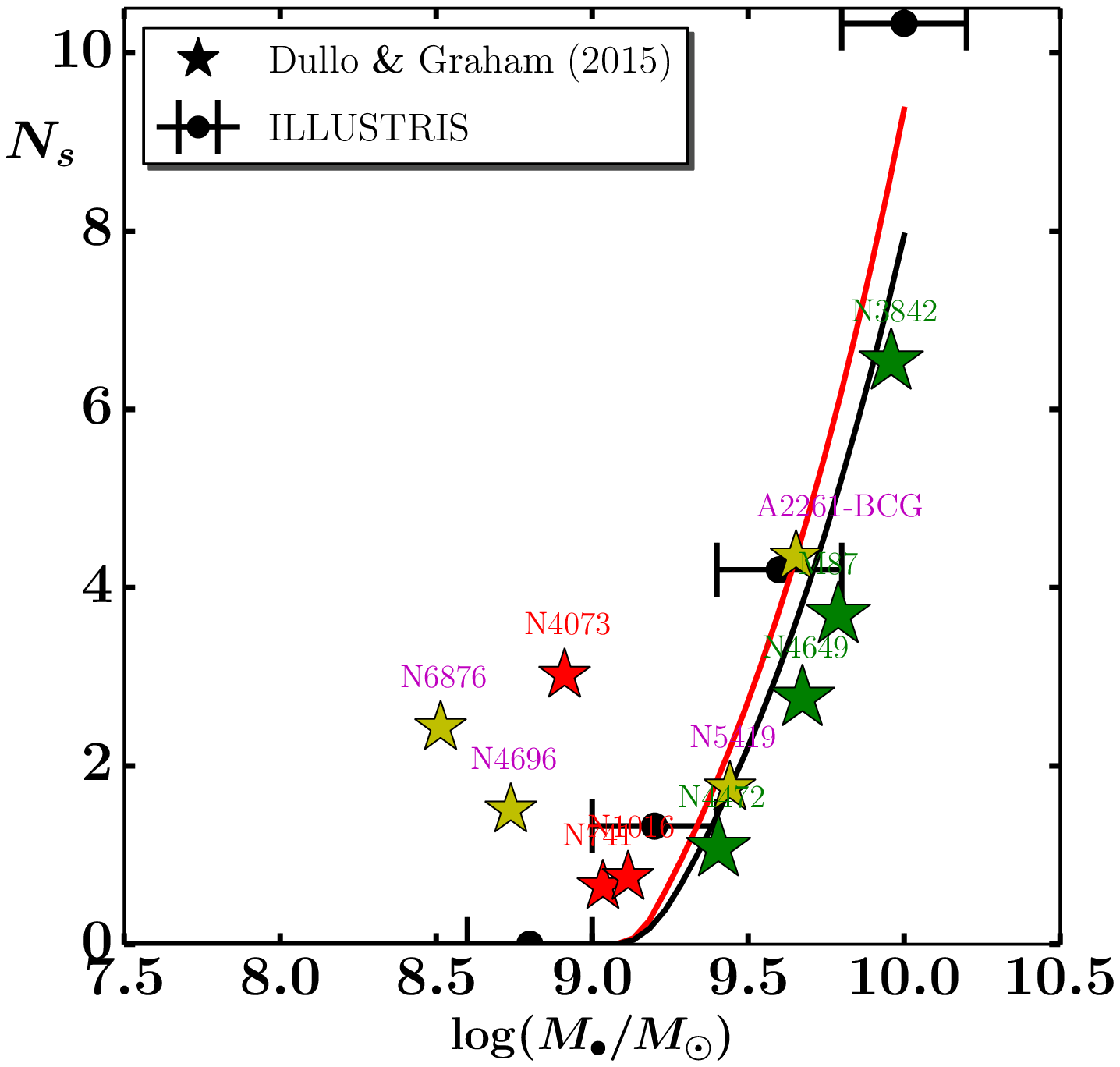}\hspace{0.2in}
   \includegraphics[angle=0,width=3.3in]{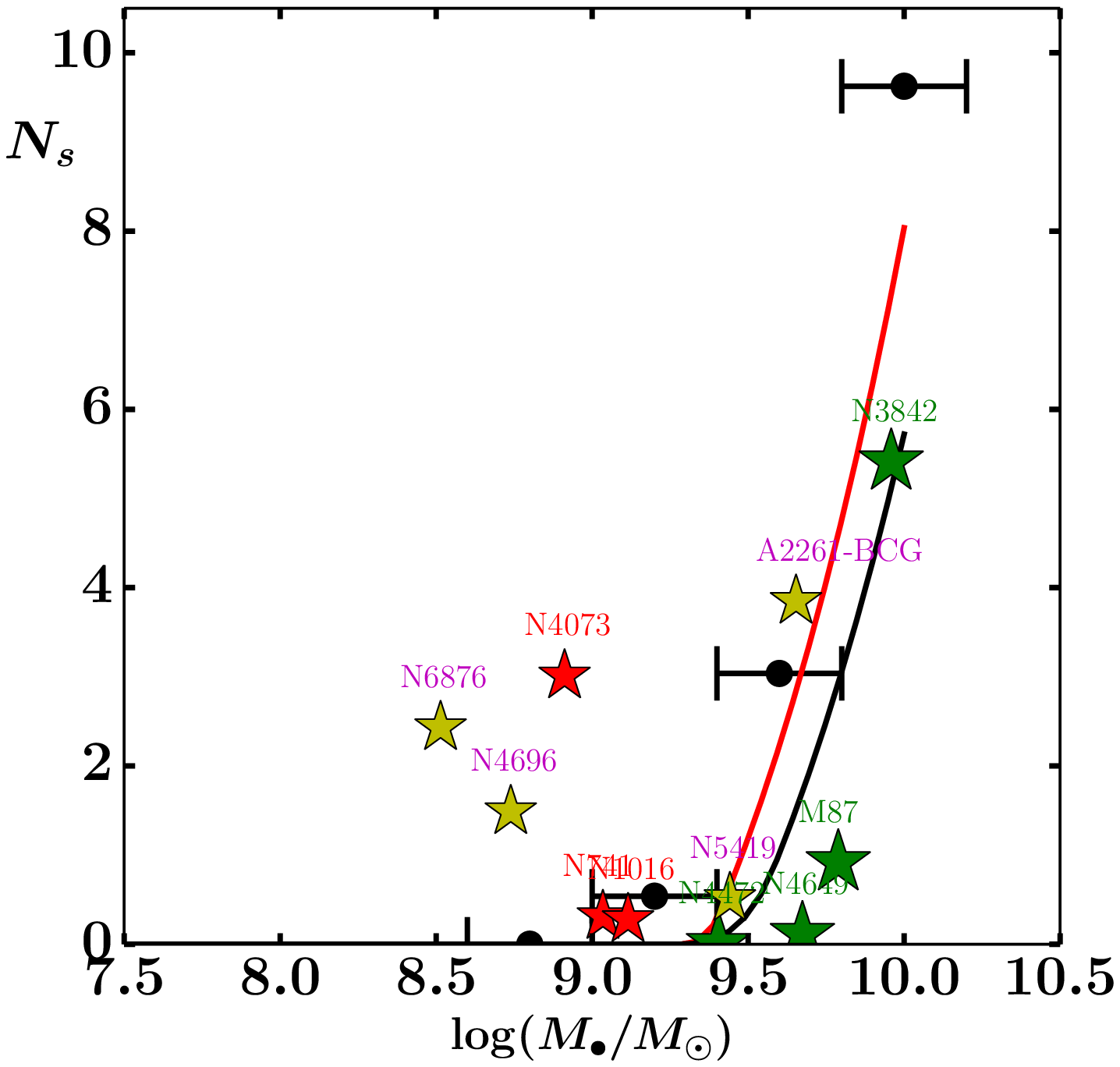}
  \caption{Expected average number of stalled satellites as a function of the host galaxy SMBH mass for $\gamma=0.6$ (left panel) and $\gamma=1.0$ (right panel). Solid red and black lines are calculated based on integral (\ref{stalled}). For the red line we used Equations (\ref{mz})-(\ref{sigmaz}) while for the black line we used, keeping the same normalization, the steepest redshift-dependence model in \citet{2012MNRAS.422.1714N}. Filled stars refer to the group of elliptical galaxies in \citet{2015ApJ...798...55D}. Galaxies with a measured primary SMBH mass are depicted as green stars, while galaxies with a SMBH mass inferred from the $M_{\bullet}-\sigma$ relation are shown as red stars. Black points are derived using the results of the ILLUSTRIS simulation \citep{2016arXiv160601900K}. The average number of stalled satellites increases similarly with the galaxy mass in all three different treatments.
The yellow stars refer to galaxies with observed double and multiple nuclei. Data for A2261-BCG, NGC 4696, NGC 5419 and NGC 6876 were taken from \citet{2012ApJ...756..159P}, \citet{2006PASA...23...33A}, \citet{2016MNRAS.462.2847M} and \citet{2002AJ....124.1975L,2012ApJ...755..163D} respectively.}
\label{stalledplot}
\end{center}
\end{figure*}
 In this calculation we
set $z_{\rm max}=1$ because at redshift lower than this ellipticals
 contain little gas and the SMBH of massive ellipticals grows primarily through minor mergers \citep[e.g.,][]{2002MNRAS.336L..61H}.
We  consider secondary SMBHs with mass $\geq 10^{6} M_{\sun}$, i.e., $q_{\rm min}=10^{6}M_{\sun}/M_{\bullet}$, central  SMBH masses  within the range $7.5<\log(M_{\bullet})<10.0$,
we use
\begin{equation}
\frac{dt}{dz}=\frac{1}{H_{0}(1+z)\sqrt{\Omega_{\rm M}(1+z)^{3}+\Omega_{\rm \Lambda}}}
\end{equation}
and assume a flat cosmology with W-MAP seven-years values for the cosmological parameters $H_{\rm 0}=70.3\: \rm (km/s)/Mpc,\Omega_{\rm \Lambda}=0.729$ and $\Omega_{\rm M}=0.1338$ \citep{2011ApJS..192...18K}.  {To give a sense of the merger inspiral efficiency we plot in Figure \ref{mergers} the average number of infalling satellites for $0<z<$1 and as a function of the host galaxy SMBH mass setting $q_{\rm crit}=1$ in Equation (\ref{stalled}).} Note that the difference $(N_m-N_s)$  is the number of BH binaries that have reached coalescence. 

 In Figure \ref{stalledplot} we plot   $N_{s}$  as a function of the host galaxy SMBH mass $M_{\bullet}$ {for $\gamma=0.6$ and $\gamma=1.0$}. The galaxy properties are related to the mass of the SMBH through Equations (\ref{mz})-(\ref{sigmaz}), while we explore the effect of a steeper redshift-dependence in the galaxy properties using a different model from \citet{2012MNRAS.422.1714N}. We find that the expected number of stalled satellites increases with the galaxy mass while adopting a different model for
 the redshift evolution of the galaxy properties does not affect significantly the number of the stalled holes.

 We do the same calculation using the observed galaxy properties for the group of elliptical galaxies in \citet{2015ApJ...798...55D} and used the redshift dependence from Equations (\ref{mz})-(\ref{rz}). For the galaxies for which we do not have a measurement of their SMBH mass we  infer this using the $M_{\bullet}-\sigma$ relation (\ref{sigmaz}) with $z=0$. We also plot the expected number of stalled satellites based on the results of the cosmological simulation ILLUSTRIS
 taken from \citet{2016arXiv160601900K}. {More specifically, for different values of redshift in the range $0<z<1$ and for the two density profiles in Figure  \ref{stalledplot} we computed the value of $q_{\rm crit}$. Using Figure 12 in \citet{2016arXiv160601900K} we calculated the number of stalled satellites we expect for a given total mass binary counting the number of mergers in the range $q<q_{\rm crit}$.  Figure 2 in \citet{2016arXiv160601900K} gives the total number of mergers in the simulation for a given total binary mass. Adding  the number of stalled satellites in the various redshift ranges and  dividing by the total number of mergers for a given total binary mass leads to an estimate for the average number of stalled satellites expected from the results in ILLUSTRIS}. Based on Figure \ref{stalledplot} the three different treatments seem to agree pretty well with each other.  For massive galaxies we expect a few stalled satellites within their inner cores.

 The question on what spatial scale the satellites stall remains still unanswered. For this purpose, we computed
  the number of black holes that stall when setting $t_{\rm L}=T_{\star}$. 
  { With this choice, for the galaxies in our sample with $10^8{M_\odot}\lesssim  M_{\bullet}\lesssim 10^9{M_\odot}$, 
  i.e., N6876, N4073 and N4696, the number of stalled satellites was roughly unchanged, demonstrating that
  in these specific galaxies the satellites are expected to reside at $\sim R_e$.
For SMBH masses larger than this, however, setting  $t_{\rm L}=T_{\star}$  gave a much smaller number
of stalled binaries in most galaxies we considered.
 We conclude that for the assumed model of hard binary inspiral in a triaxial potential,
 nearly all the SMBHs in the most massive systems we considered stall at radii  $r_{\rm h}\lesssim r \lesssim r_{\rm infl}$.}

To address more precisely where the stalling might occur we plot in Figure \ref{displacement} the time evolution of the semi-major axis
of a secondary SMBH for two mass ratios $q=(10^{-3},10^{-4})$
and for $\gamma=0.6$. The figure shows that the  orbital decay significantly
slows down at $\approx 0.1r_{\rm infl}$,
which for a typical giant elliptical galaxy corresponds to  galactocentric distances  of
tens of parsecs.

\subsection{Observational evidence}

The observational evidence of small-orbit SMBH binaries is still scarce. Electromagnetic tracers of post-merger galaxy cores are hard to identify. This makes difficult the study of the post-merger dynamics of binary SMBH systems; pairs
of SMBHs are usually observed during the early stages in their dynamical evolution when still at $\sim$ kpc separations. Although observing SMBH binaries at parsec scales is challenging since they cannot be spatially resolved in optical and X-ray, more work in detecting SMBH binaries at subkiloparsec scales ($\lesssim 100\rm \: pc$) is needed since
discovering such systems and obtaining a
census of their population properties would serve as a test of
galaxy evolution models and would provide valuable constraints
for stellar and gas dynamical models for the decay of the
binary orbit.

 If accretion is triggered along the course of the merger, direct evidence of SMBH binaries comes from the presence of active galactic nuclei (AGNs). Accretion by SMBHs can give origin to a rich phenomenology, that goes from dual to binary and off-set AGNs, in the radio-loud or quite mode, according to their dynamics, habitat and merger type. Specifically, if only one SMBH is accreting it will be visible as an off-center AGN accreting from a small disk. If both binary components of the SMBH binary accrete then a dual or double AGN might be observed.

\begin{figure}
    \includegraphics[width=.4\textwidth]{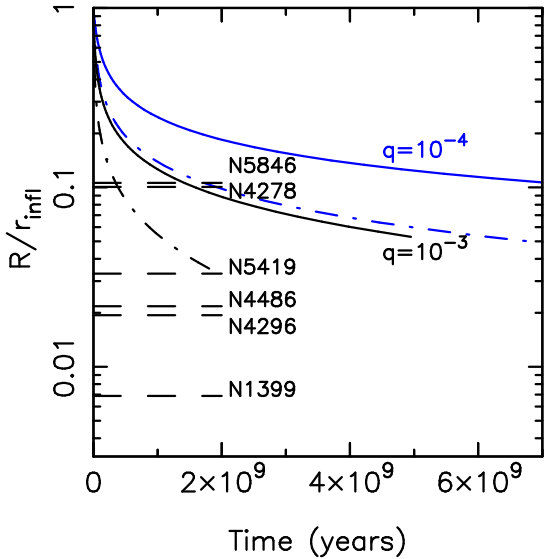}
      \caption{Evolution of orbital radius 
 for $\gamma=0.6$ (solid lines) and $\gamma=1$ (dot-dashed lines) and a binary of mass-ratio $q=10^{-4}$ (black lines) and $q=10^{-3}$ (blue lines). Dashed horizontal lines
   give the value of the observed off-center displacement reported  by \citet{2014ApJ...795..146L}.
      The satellite galaxies stall at $\approx 0.1r_{\rm infl}$.
Stalled  binaries  could  produce
 ``displacements'' comparable to those observed in NGC 4278  and NGC 5846,
if the secondary SMBH is accreting.
Note that the time has been normalized such that $M_{\bullet}=5\times 10^9 M_\odot$
and  $r_{\rm infl}=500\rm pc$, but it can be rescaled to any
of the galaxies we considered using
$t\rightarrow t \times {(r_{\rm infl}/500\rm pc)^{3/2}}(M_\bullet/5\times 10^9 M_\odot)^{-1/2}$.
}
\label{displacement}
\end{figure}

In this work we are interested in mergers that occur in gas-poor environments where collision processes are at play. Core galaxies are promising systems in which to search for SMBH binaries since they have experienced a number of
minor mergers during their lifetime and are
minimally affected by extinction.     For example, a number of displaced SMBHs has been observed as off-set AGNs in host elliptical galaxies \citep[e.g.,][]{2014ApJ...795..146L}. A characteristic example is the galaxy M87 (NGC 4486) which has a measured $7.7 \pm 0.3$ pc projected displacement of the photocenter relative to the AGN \citep{2010ApJ...717L...6B,2010ApJ...715..972J}. As a fraction of the rather large core radius $r_{\rm c}$, the weighted mean displacement is only $\approx 0.01 r_{\rm c}$.
{ The expected observed displacement $\Delta r$ of the primary SMBH is defined as $\Delta r =q\:R/(1+q)$ for  $q=10^{-4}$ and $q=10^{-3}$ with $R$ the orbital separation between the two SMBHs. Based on Figure \ref{displacement} the secondary SMBHs are expected to stall at a distance $a_{\rm stall}\approx 0.1 r_{\rm infl}$. At this radius we have $\Delta r\approx q\: a_{\rm stall}/(1+q)$ which gives $\Delta r/r_{\rm infl}\sim 10^{-4}$ $(q=10^{-3})$ and $\Delta r/r_{\rm infl} \sim 10^{-5}$ $(q=10^{-4})$. These values are small and not consistent with the observed off-center displacements mentioned in  \citet{2014ApJ...795..146L} which are of order $\sim 10^{-1}-10^{-2}$ for the massive elliptical galaxies shown in Figure \ref{displacement} following the sample of \citet{2014ApJ...795..146L}.
However, if we assume that the secondary SMBH is accreting, rather than
the primary, then the predicted offsets, in this case the galactocentric distance of the satellite SMBH, appear to be consistent at least with those observed in NGC 4278 and NGC 5846.}

Although difficult to be spatially resolved in optical and X-ray, giant ellipticals often host strong radio sources. For example, a pair of AGNs within the elliptical host galaxy 0402+379 and with a projected separation $7.2$ pc has recently been discovered \citep{2006ApJ...646...49R,2009ApJ...697...37R,2011MNRAS.410.2113B}. This is the closest binary SMBH yet discovered within a core elliptical and may be the tip of the iceberg of SMBH binaries with parsec scale separations. Binary AGNs within host core ellipticals at subkiloparsec separations like the one in \citet{2006ApJ...646...49R} could be explained as stalled holes in a slowly evolving orbit inside low-density cores.

 Early observations have suggested the presence of double nuclei in core elliptical galaxies with the second nucleus at off-center subkiloparsec separations.
 A characteristic example is the double nucleus of the core elliptical galaxy NGC 5419 with the second nucleus at an off-center separation  $\sim 70$ pc \citep{2016MNRAS.462.2847M}. The projected separation between the two nuclei is  smaller than the estimated SMBH
influence radius ($r_{\rm infl} \approx 100$ pc) and much smaller than the core
radius ($r_{\rm c} \approx 500$ pc). The same appears to be true in the case of NGC 4696 which has a secondary nucleus
at $\sim 30$ pc from the center \citep{2003AJ....125..478L}, which is slightly smaller than the estimated SMBH
influence radius ($r_{\rm infl} \approx 40$ pc) and much smaller than the core
radius ($r_{\rm c} \approx 250$ pc). These features in NGC 4696 have
been interpreted as evidence of a recent minor merger with a gas-rich
galaxy \citep{1989ApJ...345..153S,2010ApJ...724..267F}. {Similar is the case of NGC 6876 which has a double nucleus
at $\sim 30$ pc from the center \citep{2002AJ....124.1975L,2012ApJ...755..163D}. This is slightly smaller than the estimated SMBH
influence radius ($r_{\rm infl} \approx 40$ pc) and much smaller than the galaxy inner core
radius ($r_{\rm c} \approx 119$ pc).}

%A characteristic example is  the nucleus of the low-luminosity elliptical galaxy NGC 4486B (companion to M87). NGC 4486B is unusual for a galaxy of its luminosity
%in having a well-resolved core. Two
%two brightness peaks are observed in the core separated by $\approx 12$ pc  \citep{1996ApJ...471L..79L}. Neither peak is coincident with the galaxy photocenter, which falls between
%them.

More recent observations suggest the presence of multiple nuclei located (in projection) within the core elliptical galaxy A2261-BCG  \citep{2016ApJ...829...81B}. The estimated mass of the core and the central SMBH in A2261-BCG is $M_{\rm core}\sim M_{\bullet}\sim 2\times 10^{10} $.  This implies that if a SMBH is present at the center of this galaxy, then such nuclei are residing well inside its influence radius.

The double/multiple nuclei considered above have galactocentric separations
that are below $r_{\rm infl}$ but well above the separation at which the binary
can be considered a ``hard'' binary. We conclude that
the equilibrium and stability of such double/multiple nuclei  could be understood
in terms of the long dynamical friction timescale within the influence radius 
of the host galaxy SMBH predicted by our models.

We conclude by noting that the stalled satellites
predicted by our models
are likely to be on highly eccentric orbits (See Section 2).
The highly eccentric orbit of the satellite can also have interesting observational consequences. As
the satellite  is disrupted it will form an eccentric disk-like structure around the nucleus of the primary galaxy.
A characteristic example could be the nucleus of the low-luminosity elliptical galaxy NGC 4486B (companion to M87). NGC 4486B is unusual for a galaxy of its luminosity
in having a well-resolved core. Two
brightness peaks are observed in the core separated by $\approx 12$ pc  \citep{1996ApJ...471L..79L}. Neither peak is coincident with the galaxy photocenter, which falls between
them. This  double peak structure has been  interpreted as evidence for an eccentric-disk of stars
where the peaks would be the ansae of the disk  orbiting a SMBH.
The disk might be related  to the disruption 
of a star cluster on an eccentric orbit by the tidal field of the primary SMBH;
its eccentric nature would naturally result by the dynamical friction process in the flat density core.
For example,  taking $M_{\bullet}=2\times 10^8M_{\odot}$ \citep{1996ApJ...471L..79L}, an inspiraling star cluster with total mass $m=10^6M_{\odot}$ will reach the center in $\approx 0.2-0.5\rm Gyr$ starting from $r_{\rm infl}$.
If we assume a reasonable value for the cluster core radius of $\approx 2\rm pc$ then the 
cluster will be disrupted by the SMBH tidal field  
when it reaches a separation of $\approx 10\rm pc$ from the center.
This distance appears to be consistent with the observed  offsets  of the two nuclear brightness peaks  in NGC 4486B.

\section{Conclusions}

In this paper we study the orbital evolution of a 
massive binary which consists of a massive object moving near a SMBH of considerably larger mass and which sits at the center of a galaxy.
 The main physical mechanism that drives the evolution of the binary is dynamical friction.

 The main results of this paper are summarized below:
\begin{itemize}
\item[1)] We study the orbital evolution of the massive body treating dynamical friction as a perturbation to the classic two-body problem. 
Unlike previous treatments  we take into account the contribution to dynamical friction from stars moving faster than the 
massive body. Assuming that the density profile of field stars follows $\rho\propto r^{-\gamma}$,
we find that the binary secular eccentricity always increases unless $\gamma \gtrsim 2$. Specifically, low mass-ratio binaries with a moderate initial eccentricity,  $e_0\gtrsim 0.3$, attain $e\gtrsim 0.9$  by the time they reach the hardening phase. Although the contribution from fast stars increases the orbital decay rate, for cusps shallower than $\rho\propto r^{-1}$ the dynamical friction timescale becomes very long.

\item[2)] We run $N$-body simulations with different resolutions and for various slopes of the stellar cusp density profile and mass of the secondary body. We confirm the expected theoretical prediction  that for shallow density profile cusps   the eccentricity increases
 while for $\gamma\gtrsim 2$  the eccentricity decreases during the inspiral. 
\\

\item[3)] We apply our  treatment of dynamical friction to study the evolution of SMBH binaries formed in early-type galaxies. We treat independently  the different phases involved in the  evolution of the binary, compute the decay timescale that describes the dynamical friction phase and calculate the lifetime of a SMBH binary as a function of its total mass and mass ratio. We find that 
low mass ratio binaries, $q\lesssim 10^{-3}$, formed in massive elliptical galaxies  ($\gamma <1$) have a lifetime greater than a Hubble time. This results in
stalled satellite SMBHs on eccentric orbits at a galactocentric distance of order one tenth the influence radius of the primary black hole.

\item[4)] We calculate the expected  number of stalled satellites as a function of the host galaxy SMBH mass. We find that the number increases with the galaxy mass and that the brightest cluster galaxies should  have a few of such satellites. 
We discuss our  results in connection to displaced  active galactic nuclei, double and multiple nuclei often 
 observed in  core elliptical galaxies and eccentric nuclear stellar disks.

\end{itemize}

\bigskip
\section*{ACKNOWLEDGEMENTS}
We want to thank our colleague Eugene Vasiliev who provided insight and expertise that greatly assisted the research. We are
 thankful to the referee for his/her comments and suggestions that helped to improve the manuscript during its revising stages.

%\bigskip
%\section*{ACKNOWLEDGEMENTS}
%\newpage

\end{document}